\documentclass[12pt]{article}

\usepackage{graphicx}
\usepackage{amsmath}

\usepackage{amsfonts}
\usepackage{bm}
\numberwithin{equation}{section}

\def\normalorder{{\!\!\tiny \begin{array}{cc}\circ \\ \circ \\ \end{array}\!\!}} 
\def\thickone{{\rm 1\mskip-4.5mu l}}

\begin{document}

 \title{{\bf The Orbifolds of Permutation-Type as \\
 Physical String Systems\\ at Multiples~of $\mathbf{c=26}$ \\
 V.  Cyclic Permutation Orbifolds}}

\author{M.B.Halpern\footnote{halpern@physics.berkeley.edu}
\\ Department of Physics\\ University of California\\
Berkeley, Ca. 94720, USA}

\maketitle
\begin{abstract}

\noindent I consider the $\mathbb{Z}_\lambda,$ $\lambda$ prime
 free-bosonic permutation
orbifolds as interacting physical string systems at $\hat{c} = 26\lambda $. As a 
first step, I introduce twisted tree diagrams which confirm at the 
interacting level that the physical spectrum of each twisted sector is 
equivalent to that of an ordinary $c=26$ closed string. The untwisted 
sectors are surprisingly more difficult to understand, and there are
subtleties in the sewing of the loops, but I am able to propose 
provisional forms for the full modular-invariant cosmological 
constants and one-loop diagrams with insertions.

\end{abstract}

\newpage

\tableofcontents 
\newpage
\section{Introduction}

\noindent At the level of examples, current-algebraic conformal field
theory [1,2] and orbifold theory [3-8] are almost as old as string theory 
itself [9-12]. It is only in the last few years however that
the orbifold program [13-23,24-27] has in large part completed the local 
description of the general closed- and open-string orbifold conformal 
field theory. See Refs. [22,28] for short reviews of the orbifold 
program and related recent developments in orbifold theory.

Extending the results of the orbifold program, I have recently proposed 
[28-31] 
that the {\it orbifolds of permutation-type} define generically-new physical open- and 
closed-string systems at multiples of conventional critical central
charges. The simplest examples of these systems are the free-bosonic 
cases at $\hat{c}=26K$, for which we have so far studied the following
topics:

\vskip 12pt \noindent I. The extended action formulations of the twisted
 sectors of these 
orbifolds, which show {\it new permutation-twisted world-sheet 
gravities} [28]. The extended diffeomorphism invariances and extended, 
twisted Virasoro constraints of these actions indicate that, as string 
theories, the orbifolds of 
permutation-type can be free of negative-norm states. The principles 
discussed in this reference will suffice to construct the twisted world-sheet 
supergravities of the corresponding superstring orbifolds of 
permutation-type.

\vskip 12pt \noindent II. The twisted reparametrization ghosts and
 {\it new twisted BRST systems} [29] of $\mathbb{Z}_2 $-twisted permutation
 gravity, which   
imply the {\it extended physical state conditions} (extended,twisted Virasoro 
constraints) of all $\hat{c}=52$ matter. Beyond this simple case, the 
twisted BRST systems of the general world-sheet permutation gravities 
have not yet been worked out.

\vskip 12pt \noindent III. The extended Virasoro generators
 and {\it physical spectra} 
of all twisted $\hat{c}=52$ strings  [30], including an equivalent, 
unconventionally-twisted $c=26$ 
description of their spectra. This analysis provides further evidence 
that the orbifold-string systems of permutation-type can be free of negative-norm
states, and moreover shows that a few of the simplest (half-integer 
moded) open and closed
$\hat{c}=52$ strings have the same spectra as ordinary untwisted $c=26$
strings. Beyond these simple cases, the twisted $\hat{c}=52$ strings are 
apparently new.

\vskip 12pt \noindent IV. {\it Orientation orbifolds include orientifolds}  
[27].
 The orientation-orbifold string systems [28-30] involve dividing the closed string by 
 automorphism groups which contain world-sheet orientation-reversing 
 automorphisms. They therefore contain an equal number of generically-twisted
closed- and open-string sectors, each of the latter at $\hat{c}=52$. 
In this fourth paper of the series, twisted tree diagrams are constructed to study a 
particular example with a single half-integer moded open-string 
sector, and evaluation of the trees confirms [30] at the interacting level
that this orientation-orbifold system is nothing but the archetypal 
orientifold in disguise. There are many other free-bosonic orientation
orbifolds (with higher fractional modeing) which are not equivalent 
[29] to conventional orientifolds, but these ``generalized 
orientifolds'' have not yet been studied at 
the interacting level.

In this fifth paper of the series, I return to the closed-string 
permutation orbifolds at central charge $\hat{c}=26K$, whose general form

\begin{equation} 
\frac{U(1)^{26K}}{H_{+}},\quad 
 H_{+}=H(\text{perm})\times H,\,\,\,
 H(\text{perm})\subset S_{K}
\end{equation}

\noindent includes the uniform action of automorphisms $H$ on each
 untwisted closed-string copy $U(1)^{26}$. For simplicity, the discussion of 
 these orbifolds is 
 continued here only for the case of continuous (decompactified) 
 zero modes, but this can be straightforwardly generalized.
The physical spectrum of all the twisted sectors at $\hat{c}=52$ (all $H_{+}$ 
with $K=2$ and $H(\text{perm})=\mathbb{Z}_2$) was studied in Ref. [30], 
where it was noted in particular that the spectrum of the twisted sector with 
trivial $H$ (the ordinary $\mathbb{Z}_2$-permutation orbifold)  is equivalent
to that of an ordinary untwisted $c=26$ closed 
string. Generalizing the $\mathbb{Z}_2$ orbifold, I will restrict
 the discussion here to the string theories of the simplest {\it cyclic 
 permutation orbifolds}

\begin{equation} 
\frac{U(1)^{26\lambda}}{\mathbb{Z}_\lambda},\quad \lambda\, 
\text{prime},\,\,\, 
\hat{c}=26\lambda
\end{equation}

\noindent but I will consider this family at the {\it interacting 
level} through one loop.
 
 As a first step, I introduce twisted tree diagrams (see Sec. 2) for 
 these orbifolds which 
 confirm at the interacting level for all prime $\lambda$ our earlier 
 conclusion [30] for the free theory at $\lambda =2$: The physical 
 spectrum of each of the $\lambda -1$ twisted sectors of 
 $U(1)^{26\lambda}/\mathbb{Z}_{\lambda}$ is equivalent to 
 that of an ordinary untwisted $c=26$ string. As emphasized in Refs. 
 [30,31], this so-called
 {\it spectral equivalence} holds {\it only} for the string amplitudes, 
 which involve extra integrations over the correlators of the orbifold
 CFT's.
 
 Sewing the twisted trees gives an integrated form of the standard 
 [7]
 twisted contribution to the orbifold partition function, and the 
 following formal integration identity (see Eq. 3.10)
 
\begin{subequations} 
\begin{gather} 
\int_{ F_{N}} \frac{d^{2}\tau}{({\rm Im}\tau)^{2}} \left( Z(\tau, \bar{\tau}) 
- \sum_{r=0}^{\lambda -1} 
 Z\left(\tfrac{\tau + r}{\lambda}, \tfrac{\bar{\tau} 
 +r}{\lambda}\right)\right) = 0 
 \end{gather}
 \begin{gather} 
  F_{N}:\,\,\, |{\rm Re}\tau|\leq \frac{1}{2}, \,\, {\rm Im}\tau\geq 0 
\end{gather}
\end{subequations}
  
 \noindent then verifies the same spectral equivalence on the torus. 
 Here  $Z(\tau, 
 \bar{\tau})$ is the partition function of the ordinary untwisted closed 
 string $U(1)^{26}$, and $F_{N}$ is the {\it naive} integration range encountered in the 
 {\it naive (operator) sewing} [31,9] of any string loop.  This 
  operator sewing is discussed first for the 
 orbifold cosmological constant in Subsecs 3.2 and 3.3, where it is called 
 summing over the {\it particle-theoretic content} of the theory,
  and later for the loop with insertions in Sec. 6.
 
 The particle-theoretic contributions of the untwisted sector are
  surprisingly more 
 difficult to understand: I had expected to find included here the non-linear 
 untwisted contribution $Z^{\lambda}(\tau,\bar{\tau})$ of the standard 
 orbifold partition function [6], but in fact (see Subsec. 4.1) any such
 non-linear contribution is incompatible with one-loop structure in string 
 theory!
  
 Indeed I argue in Subsec. 4.2 and Sec. 6 that the particle-theoretic
  contribution of 
 the untwisted sector to the sewn loop is equivalent to that of 
 $\lambda$ ordinary untwisted closed strings, which is also consistent with the
 cross-channel behavior of the twisted trees. Moreover, according to 
 a second formal integration identity (see Eq. (4.8))
 \begin{equation} 
\int_{ F_{N}} \frac{d^{2}\tau}{({\rm Im}\tau)^{2}} \Big{(}
\lambda Z(\tau,\bar{\tau}) - Z(\lambda\tau, \lambda\bar{\tau})\Big{)} = 0
\end{equation}

\noindent this situation can be equivalently described by keeping 
only the so-called {\it diagonal contribution} $Z(\lambda\tau, 
\lambda\bar{\tau})$ to the untwisted sector (see Refs. [6] and 
[9]).

Secs. 5 and 6 finally assemble our observations about the twisted and 
untwisted loop contributions. The central point here is that the 
formal integration identities (1.3) and (1.4) generate certain {\it ambiguities}
in the naive or particle-theoretic form of the loops, which persist in the 
corresponding modular-invariant forms. Thus I am able to propose for 
future study the following one-parameter {\it $\beta$-family of provisional forms} for 
the full modular-invariant cosmological constants (see Eq. (5.1))

\begin{subequations} 
\begin{gather} 
\begin{split}
(\alpha_{c}^{'})^{13}\Lambda^{(\lambda)}(\beta) =
  -\frac{1}{2}\int_{ F} \frac{d^{2}\tau}{({\rm Im}\tau)^{2}} &\Big{\{}
  (2\lambda -1 -(\lambda + 1)\beta) Z(\tau, \bar{\tau})\,+
\\ 
&+\beta \left\lbrack Z(\lambda\tau,\lambda\bar{\tau})
 + \sum_{r=0}^{\lambda -1}Z\left(\tfrac{\tau +r} {\lambda},
  \tfrac{\bar{\tau} +r}{\lambda}\right)\right\rbrack\Big{\}}
\end{split}
\end{gather}
\begin{gather} 
  F:\,\,\, |{\rm Re}\tau|\leq \frac{1}{2}, \,\, {\rm Im}\tau\geq 0, \,\,|\tau|\geq 1
\end{gather}
\end{subequations}

\noindent where F is the standard fundamental region of the modular 
group. In these provisional forms, the parameter $\beta$ describes the ambiguity in the naive 
sewing of the loops. The corresponding results for the one-loop diagrams with insertions 
are given in Eq. (6.1).

I emphasize that the simplest member of this family is $\beta=0$
\begin{equation} 
(\alpha_{c}^{'})^{13} \Lambda^{(\lambda)}=
-\frac{2\lambda -1}{2}\int_{ F_{N}} \frac{d^{2}\tau}{({\rm Im}\tau)^{2}}
\,Z(\tau,\bar{\tau})
\end{equation}

\noindent which is clearly equivalent to the contribution of $\lambda$ ordinary $c=26$ strings from the untwisted sector 
and $\lambda -1$ ordinary $c=26$ strings from the twisted sectors. 
Other interesting special cases are discussed in Subsec. 5.2 but, since all
values of $\beta$ are formally associated to the {\it same} naive sewing of the 
trees, the entire $\beta$-family should be further tested against other 
string- and conformal-field-theoretic intuitions.

In this connection, Subsec. 5.2  also includes some remarks (see in 
particular  Eq. (5.9)) on 
hitherto-unnoticed 
{\it modular-covariant subsets} of permutation-orbifold characters -- which 
exclude the so-called off-diagonal characters [10] of the untwisted 
sectors. These closed subsets are the natural infrastructure for 
partition functions of the form (1.5) with no non-linear term 
$Z^{\lambda}(\tau, \bar{\tau})$, and they may be useful in deciding 
among the provisional forms.

\newpage

\section{The Twisted Sectors on the Sphere}

\subsection{The Twisted Algebras of Sector $\mbox{\boldmath{$\sigma$}}$}

\noindent Before considering their corresponding string theories, I 
collect here some useful facts about our class of orbifolds as 
{\em conformal field theories}. 

The cyclic permutation orbifold [13-15,17-19,21]
\begin{equation} 
 \frac{U(1)^{26\lambda}}{\mathbb{Z}_\lambda},\,\,\,\, \lambda\,
 \text{prime},\,\,\,\hat{c}=26\lambda
\end{equation}
 \noindent  has $\lambda -1$ identical twisted sectors $\sigma = 
1,\ldots,\lambda -1$ at central charge $26\lambda$. In each twisted sector, one finds the 
$\sigma$-independent (left-mover) extended Virasoro generators and twisted 
algebras [13]:
\begin{subequations} 
\begin{gather} 
\begin{split} 
\hat{L}_{r}(m + \tfrac{r}{\lambda}) =
&\,\,\frac{13}{12}\left( \lambda -\frac{1}{\lambda} \right) \delta_{m+\tfrac{r}{\lambda},0}\\
&-\frac{\eta^{\mu\nu}}{2\lambda} \sum_{s=0}^{\lambda -1} \sum_{p}
\normalorder\hat{J}_{s\mu}(p+\tfrac{s}{\lambda})\hat{J}_{r-s,\nu}(m-p+\tfrac{r-s}{\lambda})
\normalorder_{M}
\end{split} 
\end{gather}
\begin{gather} 
\lbrack\hat{J}_{r\mu}(m+\tfrac{r}{\lambda}),
\hat{J}_{s\nu}( n+\tfrac{s}{\lambda})\rbrack =
-\lambda\eta_{\mu\nu}(m+\tfrac{r}{\lambda})
\delta_{m+n+\tfrac{r+s}{\lambda}, 0}
\end{gather}
\begin{gather}
\lbrack\hat{L}_{r}(m+\tfrac{r}{\lambda}),\hat{J}_{s\mu}(n+\tfrac{s}{\lambda}\rbrack =
-(n+\tfrac{s}{\lambda})\hat{J}_{r+s,\mu}(m+n+\tfrac{r+s}{\lambda})
\end{gather}
\begin{gather}
\begin{split} 
\lbrack\hat{L}_{r}(m+\tfrac{r}{\lambda}), 
       \hat{L}_{s}(n+\tfrac{s}{\lambda})\rbrack =
&(m-n +\tfrac{r-s}{\lambda}) \hat{L}_{r+s}(m+n +\tfrac{r+s}{\lambda})\\
&+\frac{26\lambda}{12} 
(m+\tfrac{r}{\lambda})((m+\tfrac{r}{\lambda})^2
-1)\delta_{m+n+\tfrac{r+s}{\lambda},0}
\end{split}
\end{gather}
\begin{gather} 
\eta =
\left(\begin{array}{cccc}
1 & 0 \\
0 & -\thickone 
\end{array}\right), \quad\, \mu,\nu = 0,1,\ldots25, \quad \bar{r},\bar{s}=
0,1,\ldots\lambda -1.
\end{gather}
\end{subequations}
 Here $\eta$ is the Minkowskian target-space metric of the 
ordinary critical closed string $U(1)^{26}$,  $\normalorder \, \cdot \, \normalorder_{M}$ is standard mode
normal ordering [13-15,17-19,21] and Eq. (2.2d) is an orbifold Virasoro algebra 
[13,21,29,32] of order 
$\lambda$. The physical Virasoro generators $\{\hat L_{0}(m)\}$ of 
each sector satisfy the implied integral Virasoro subalgebra at $\hat 
c=26\lambda$.

Including the twisted vertex operators 
$\hat{g}(\mathcal{T})=\,:\!exp(i\mathcal{T}\!\cdot \hat{x})\!:$ of 
sector $\sigma$, one 
finds the additional algebras [15,16,18] :
\begin{subequations} 
\begin{gather} 
\lbrack\hat{J}_{r\mu}(m+\tfrac{r}{\lambda}), 
\hat{g}(\mathcal{T},\bar z, z)\rbrack = 
t_{r}T_{\mu}z^{m+\tfrac{r}{\lambda}} \hat{g}(\mathcal{T} ,\bar{z},z)
\end{gather}
\begin{gather} 
\lbrack \hat{L}_{r}(m+\tfrac{r}{\lambda}),  \hat{g}(\mathcal{T} 
,\bar{z},z)\rbrack = t_{r}z^{m+\tfrac{r}{\lambda}}(z\partial_z
+(m+\tfrac{r}{\lambda}+1))
\Delta(T)\hat{g}(\mathcal{T} ,\bar{z},z)
\end{gather}
\begin{gather} 
\lbrack T_{\mu},\hat{g}(\mathcal{T} ,\bar{z},z)\rbrack = \lbrack t_{r},
 \hat{g}(\mathcal{T} ,\bar{z},z)\rbrack=0
 \end{gather}
 \begin{gather} 
 \mathcal{T}_{r\mu}=t_{r}T_{\mu},\quad \lbrack T_{\mu},T_{\nu}\rbrack 
 =0,\quad \Delta(T)= -\frac{T^{2}}{2}
 \end{gather}
 \begin{gather} 
 (t_{r})_{s}^{\ t}= \delta_{r+s-t,o{\rm mod}\lambda},\quad t_{r}t_{s}= 
 t_{r+s},\quad t_{0}=\thickone_{\lambda}.
 \end{gather}
 \end{subequations}
 In these relations,  $\{T\}$ are the dimensionless momenta (for 
 simplicity assumed here to be continuous) and $\hat{g}$ has the 
 same $\lambda\times\lambda$ matrix 
 structure as the matrices $\{t\}$. There are also right-mover copies 
 $\{\hat{\bar{J}}_{r\mu}\}$ and $\{\hat{\bar{L}}_{r}\}$ in each twisted sector with the same 
 algebra (2.2) and $z\rightarrow \bar{z}$ in the coefficients of Eqs. 
 (2.3a,b). The explicit form
 and diagonal monodromy of the twisted vertex operators 
 \footnote{For simplicity I have here taken 
 $T\rightarrow -T$ in the right-mover vertex operators of 
 Ref. [19], which uniformizes the right- and left- mover signs in 
 Eqs. (2.3a,b) without changing any orbifold correlators.} of all free-bosonic 
 orbifolds are given in Ref. [16], 
 and a concrete  example will be considered in Subsec. 2.4.
 
 We will also need the momentum-boosted twist-field state 
 $|\mathcal{T}\rangle$ of sector $\sigma$ which satisfies [18,19,21]
 \begin{subequations} 
\begin{gather} 
|\mathcal{T}\rangle = |T\rangle \thickone_{\lambda} = \lim_{z \rightarrow 0} 
 |z|^{2\Delta(T)(1-\lambda^{-1})} \hat{g}(\mathcal{T} ,\bar{z},z) 
|0\rangle
\end{gather}
\begin{gather} 
\hat{J}_{r\mu}((m+\tfrac{r}{\lambda}) \geq 0) |T\rangle = 
\delta_{m+\tfrac{r}{\lambda},0} \, |T\rangle T_{\mu}
\end{gather}
\begin{gather} 
\hat{L}_{r}((m+\tfrac{r}{\lambda}) \geq 0)  |T\rangle = 
\delta_{m+\tfrac{r}{\lambda},0}  |T\rangle 
\left(\frac{13}{12}\left(\lambda - \frac{1}{\lambda}\right) + \frac{\Delta(T)}{\lambda}\right)
\end{gather}
\end{subequations}
 as well as right-mover copies of Eqs. (2.4b,c). Note that 
the state $|T\rangle$ in these relations is a state in the oscillator 
Hilbert space, while the state $|\mathcal{T}\rangle$ has an 
additional (trivial) matrix structure.
 
\subsection{The Twisted Trees of Sector $\mbox{\boldmath{$\sigma$}}$}
 
\noindent I begin the construction of the permutation-orbifold {\em string 
systems} by defining
 the {\it string propagator} in twisted sector $\sigma$:
 \begin{subequations} 
 \begin{gather} 
 \hat{D} \lbrack \hat{L}_{0}(0), \hat{\bar{L}}_{0}(0)\rbrack  \equiv \,\, 
 \frac{{\delta _{{\hat L}_0 (0),{\hat {\bar L}}_0 (0)} }}
{{\lambda ({\hat L}_0 (0) + {\hat {\bar L}}_0 (0) - 2{\hat a}_\lambda  )}}
 \end{gather}
 \begin{gather} 
 \hat{a}_{\lambda} \equiv \,\, \frac{13\lambda^{2}-1}{12\lambda}.
 \end{gather}
 \end{subequations}
 The factor in the numerator of $\hat{D}$ is ordinary Kronecker 
 delta and, according to the twisted BRST system of Ref. [29], the quantity 
 $\hat{a}_{2} = 17/8$ is universal for all $\hat{c}=52$ matter. The form of 
 $\hat{a}_{\lambda}$ for $\lambda\geq 3$ was conjectured in Ref. [30], 
 and we shall check the consistency of this conjecture in the following
 subsection. Useful integral representations of the propagator include
\begin{equation} 
\begin{split}
  {\hat D}[{\hat L}_0 (0),{\hat {\bar L}}_0 (0)]
  & = \int_{|z| \leq 1} {\frac{{d^2 z}}
{{2\pi \lambda }}z^{{\hat L}_0 (0)} {\bar z}^{{\hat {\bar L}}_0 (0)} |z|^{ - \frac{{(13\lambda  - 1)(\lambda  + 1)}}
{{6\lambda }}} \frac{1}
{\lambda }\sum\limits_{r = 0}^{\lambda  - 1} {e^{2\pi ir\left( {{\hat L}_0 (0) - {\hat {\bar L}}_0 (0)} \right)} } }  \hfill \\
   =& \int_{|z| \geq 1} {\frac{{d^2 z}}
{{2\pi \lambda }}z^{ - {\hat L}_0 (0)} {\bar z}^{ - {\hat {\bar L}}_0 (0)} |z|^{\frac{{(13\lambda  + 1)(\lambda  - 1)}}
{{6\lambda }}} \frac{1}
{\lambda }\sum\limits_{r = 0}^{\lambda  - 1} {e^{ - 2\pi ir\left( {{\hat L}_0 (0) - {\hat {\bar L}}_0 (0)} \right)} } }  \hfill \\ 
\end{split} 
\end{equation}
where the sums set ${\hat {\bar L}}_0 (0)\simeq \hat{L}_{0}(0)$ modulo 
the integers.


The conformal weight of the orbifold-string ground state $|T\rangle$ and its 
twisted vertex operator $\left.\hat{g}\right.(\mathcal{T})$ is then determined
\begin{subequations} 
\begin{gather} 
\Delta(T) = \lambda \left( {{\hat a}_\lambda   - \frac{{13}}
{{12}}\left( {\lambda  - \frac{1}
{\lambda }} \right)} \right) = 1,\quad T^2  =  - 2 \\
\left( {{\hat L}_r \left( {\left( {m + \tfrac{r} 
{\lambda }} \right) \geq 0} \right) - {\hat a}_\lambda \, \delta _{m + \frac{r}
{\lambda },0} } \right) |T\rangle = 0
\end{gather}
\end{subequations}
 by comparing the poles of the propagator to Eq. (2.4c). In this case, 
the vertex operator ``boost'' relation
\begin{equation} 
{\hat g}(\mathcal{T},{\bar z},z) = z^{{\hat L}_0 (0)} {\bar z}^{{\hat {\bar L}}_0 (0)} {\hat g}(\mathcal{T},1,1)z^{ - {\hat L}_0 (0) - 1} {\bar z}^{ - {\hat {\bar L}}_0 (0) - 1} 
\end{equation}

\noindent follows from Eq. (2.3b).


We are now ready to define the ($\sigma$-independent) n-point {\it twisted 
tree graphs} of each twisted sector $\sigma$ of the orbifold-string 
system  by the so-called sidewise construction [33,3,12,31]
\begin{subequations} 
\begin{gather} 
\begin{split}
  {\hat A}_n^{(\lambda )} (\{ \mathcal{T}\} ) \equiv \, S^T \langle  - T^{(n)} | &{\hat g}(\mathcal{T}^{(n - 1)} ,1,1){\hat D}[{\hat L}_0 (0),{\hat {\bar L}}_0 (0)]{\hat g}(\mathcal{T}^{(n - 2)} ,1,1) \hfill \\
  & \cdots {\hat D}[{\hat L}_0 (0),{\hat {\bar L}}_0 (0)]{\hat g}(\mathcal{T}^{(2)} ,1,1)|T^{(1)} \rangle S +  \cdots  \hfill \\  
\end{split}\\ 
S_r  = \left( {S^T } \right)^r  = \frac{1}
{{\sqrt \lambda  }},\quad\,\, {\bar r} = 0,1, \ldots ,\lambda  - 1, \quad T^{2}=-2
\end{gather}
\end{subequations}
which shows any particular twisted $\hat c=26\lambda$ sector running 
horizontally (sidewise). 
Here the product of the twisted vertex operators is defined by matrix 
multiplication
\footnote{This choice seems necessary to obtain the 
extended Ward identities of the following subsection.}
 with all $\{t_{r}^{i}=t_{r}\}$ , and the final ellipsis 
denotes symmetrization with respect to the arguments of the vertex 
operators. Finally, note that the normalized ``spinor'' $S$ 
 in Eq. 2.9b is the 
simultaneous eigenvector of the $t$ matrices
\begin{equation} 
S^TS=1,\quad t_r S = \thickone_\lambda  S,\quad S^T t_r  = S^T 
\thickone_\lambda  ,\quad\,\, {\bar r} = 0, \ldots ,\lambda  - 1
\end{equation}
where I remind that $t_{0}=\thickone_\lambda$ is the $\lambda \times 
\lambda$ unit matrix (see Eq. (2.3e)). This completes my definition of
the twisted trees, whose properties will now be studied in some detail. 


I turn first to some properties of the external particles emitted from 
the twisted sectors (channels) of these tree graphs, which will allow 
us to understand the hidden cross channel of the twisted trees as 
the {\it untwisted channel} of each cyclic permutation orbifold.

In this connection, note that 
the properties (2.10) of the spinor $S$ allow us to write the  vertex operator on $S$ in a variety of ways
\begin{subequations} 
\begin{gather} 
\hat g(\mathcal{T}) S\,\,=\,\,   \hat{\hat g}(T) S\,\,=\,\,\hat g_I(T)S, 
\quad\,\, I=0,1,\ldots,\lambda-1\\ 
\hat g(\mathcal{T}) \sim \normalorder \exp(i\sum_{r \mu} t_r T_\mu 
\hat{x}_{r\mu})\normalorder\\ 
\hat{\hat g}_I(T)   \sim \normalorder \exp(it_I\sum_{r\mu}T_\mu\hat 
{x}_{r\mu}) \normalorder\\ 
\thickone_\lambda\hat g_0(T)\,\, =\,\, \thickone_\lambda\hat{\hat g}(T)\\ 
\hat{\hat g}(T)\,\, =\,\,\hat g(\mathcal{T} \rightarrow T) \sim \normalorder 
\exp(i \sum_{r\mu} T_\mu  \hat{x}_{r\mu})\normalorder 
\end{gather}
\end{subequations}
and these properties imply the following identities on the twisted 
trees:


\begin{subequations} 
\begin{align}
\hat A^{(\lambda)}_n(\{\mathcal{T}\}) &= {\hat A}^{(\lambda)}_n(\{ T\})_I, \quad I=0, 1, \ldots, \lambda -1 \\ 
  & \equiv S^T\langle -T^{(n)} |\, \hat g_I \hat D \hat g_I \ldots \hat 
  g_I \hat D \hat g_I \,| T^{(1)}\rangle S\,+\ldots \\ 
  & = \langle -T^{(n)} |\, \hat{\hat g} \hat D \hat{\hat g} \ldots \hat{\hat 
  g} \hat D \hat{\hat g}\, | T^{(1)} \rangle\,+\ldots\,\,\,\,. 
\end{align}
\end{subequations}
The equivalences (2.11) and (2.12) allow us to interpret the vertex operator 
$\hat{g}_{I}$ as the {\em emission vertex} for a tachyon $(T^{2}=-2)$ of 
type $I= 0,1,\ldots, \lambda-1$ from each twisted channel $\sigma= 
1,\ldots,\lambda-1$. Each of the tachyon types is in fact spinless and 
untwisted because the amplitude $\hat{A}_{I}=\hat A$ 
for emission of many tachyons of type $I$ has no 
target-space or world-sheet indices. The $I$-independence $\hat{A}_{I}=\hat A$ 
of the multiple-emission amplitudes also tells us that all $\lambda$ of these tachyons are identical 
because  they couple identically to each of the $\lambda-1$ twisted 
channels.  This set of $\lambda$ identical untwisted 
tachyons is then exactly what is expected (before symmetrization on 
$\{I\}$) in the {\it untwisted sector} (cross channel) of the orbifold 
$ U(1)^{26\lambda}/\mathbb{Z}_{\lambda}$. These remarks are illustrated in Fig. 1,
where the external untwisted 
tachyon lines are dashed and the solid line is any particular twisted 
sector of the orbifold. 

\begin{figure}[htbp]
	\centering
		\includegraphics[angle=0,width=0.5\columnwidth]{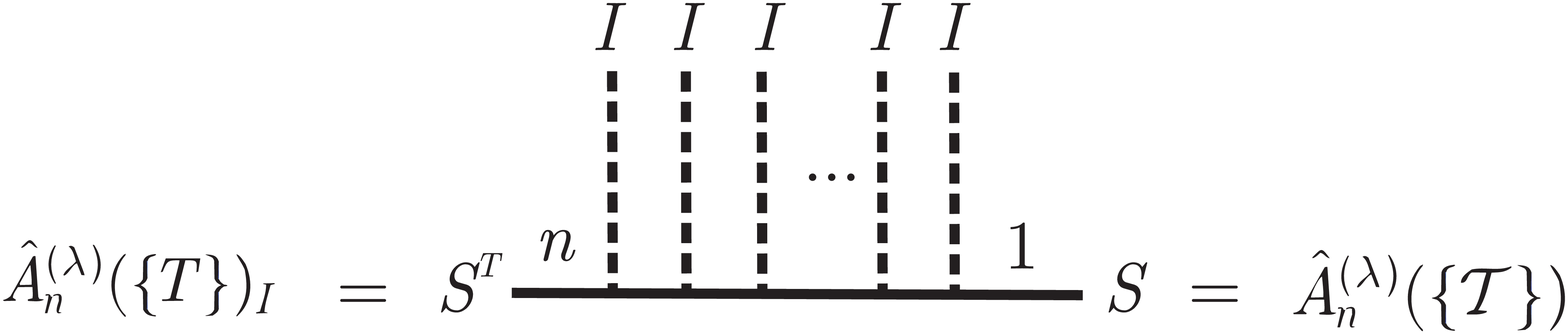}
	\caption{The twisted trees of type I are independent of I.}
	\label{fig:regionsF}
\end{figure}


There are other useful forms of the twisted trees. For example, we 
may use the integral representations in Eq. (2.6) and the boost relation 
in Eq. (2.8) to put the four-point amplitudes in 
the form
\begin{multline} 
{\hat A}_4^{(\lambda )} (\{ \mathcal{T}\} ) = \int \frac{{d^2 z}}
{{2\pi \lambda ^2 }}\sum\limits_{r = 0}^{\lambda  - 1} \times \\ 
\times S^T \langle  - T^{(4)} | R(\hat g(\mathcal{T}^{(3)} ,1,1) \hat g(\mathcal{T}^{(2)} ,{\bar z}e^{ - 2\pi ir} ,ze^{2\pi ir}) | \mathcal{T}^{(1)} \rangle S
\end{multline}
where R is radial ordering and $z$ is now integrated over the full complex plane. Note that the integrands here are explicitly 
monodromy-invariant $ (z\rightarrow ze^{2\pi i}) $ as they should be. 
This formula can be straightforwardly generalized to all the n-point 
amplitudes, and each of these can be further simplified by the substitution  
$ \hat{g}\rightarrow \hat{\hat g},\, S\rightarrow 1 $ as above.

\subsection{Extended Ward Identities and Physical States}

Our next topic is the physical spectrum of the twisted 
orbifold-string sectors, as implied by the twisted trees.

For each twisted sector $\sigma$, I define the extended (left-mover) 
gauge operators
\begin{subequations} 
\begin{gather} 
{\hat W}_r \left( {m + \tfrac{r}{\lambda }} \right) \equiv\, t_{ - r} {\hat L}_r \left( {m + \tfrac{r}
{\lambda }} \right) - \left( {{\hat L}_0 (m) + \left( {m + \tfrac{r}
{\lambda }} \right) - {\hat a}_\lambda  } \right)\\ 
\bar r = 0, \ldots ,\lambda  - 1 
\end{gather}
\end{subequations}
and right-mover copies with $\hat{L}\rightarrow \hat{\bar L}$. Useful 
identities for the study of these gauges include:
\begin{subequations} 
\begin{gather} 
  [ {t_{ - r} {\hat L}_r \left( {m + \tfrac{r}
{\lambda }} \right) - {\hat L}_0 (m),{\hat g}(\mathcal{T}^{(i)} ,1,1)} ]\, =\, \left( {m + \tfrac{r}
{\lambda }} \right){\hat g}(\mathcal{T}^{(i)} ,1,1) \hfill \\
  {\hat W}_r ( {m + \tfrac{r}
{\lambda }} ){\hat g}(\mathcal{T}^{(i)} ,1,1)\, =\, {\hat g}(\mathcal{T}^{(i)} ,1,1)\left( {t_{ - r} {\hat L}_r \left( {m + \tfrac{r}
{\lambda }} \right) - {\hat L}_0 (0) + {\hat a}_\lambda  } \right) \hfill \\ 
\begin{split}
\left( {t_{ - r} {\hat L}_r \left( {m + \tfrac{r}
{\lambda }} \right) - {\hat L}_0 (0) + {\hat a}_\lambda  } \right)
\hat D[{\hat L}_0 (0),{\hat {\bar L}}_0 (0)]  \\ 
 \quad\simeq {\hat D}[ {{\hat L}_0 (0) + m + \tfrac{r}
{\lambda },{\hat {\bar L}}_0 (0)} ]{\hat W}_r \left( {m + \tfrac{r}
{\lambda }} \right). 
\end{split}
\end{gather} 
\end{subequations}
Here Eq. (2.15a) is an orbifold generalization of the so-called  "stability 
condition" of Refs. [1,31], and Eq. (2.15b) follows from (2.15a). The 
proof of Eq. 
(2.15c) requires the extended Virasoro algebra (2.2d), and the symbol $ \simeq $ 
means that I have also used the "cancelled propagator" argument (see e.g. 
Ref. [12]).


Then we see that the extended gauges are operative in the twisted 
trees\footnote{Although there is no need to do so here, we know from 
Subsec. 2.2 that the replacement $ t_{r}\rightarrow 
\thickone_{\lambda},\, \hat{g}\rightarrow \hat{\hat g} $ can be made in 
all the operators of Eq. (2.16) on $ S $.}
\begin{equation} 
{\hat W}_r \left( {\left( {m + \tfrac{r}
{\lambda }} \right) > 0} \right){\hat g}(\mathcal{T}^{(m)} ,1,1){\hat D} \cdots {\hat D}{\hat g}(\mathcal{T}^{(1)} ,1,1)\left| \chi  \right\rangle S =\, 0
\end{equation}
so long as the physical oscillator states $\{|\chi\rangle\} $ satisfy the 
{\it extended physical state condition}:
\begin{equation} 
\left( {{\hat L}_r \left( {\left( {m + \tfrac{r}
{\lambda }} \right) \geq 0} \right) - {\hat a}_\lambda  \delta _{m + \frac{r}
{\lambda },0} } \right)\left| \chi  \right\rangle  = 0,\quad\,\, \bar r = 0, \ldots ,\lambda  - 1.\end{equation}
This condition (and a right-mover copy on the same states $\{| \chi \rangle\}$) is the operator realization 
of the extended classical Virasoro constraints of 
$\mathbb{Z}_\lambda $-twisted permutation gravity [28].

Of central importance, the 
case $ \lambda=2 $ of the extended physical state condition (2.17) is in precise agreement with that 
obtained from the twisted BRST system of $\hat{c}=52$ 
matter in Ref. [29]. Moreover, exactly this form of the extended physical state 
condition was conjectured for $\lambda\geq 3$ in Ref. [27]. Note 
finally that the ground state $ |T\rangle $ of the twisted sector in Eq. (2.7b) is 
indeed a physical state for each $\lambda$. 


What is perhaps surprising is that, {\it as a string theory}, the physical 
spectrum of each twisted $ \hat{c}=26\lambda $ sector is in fact the same 
as that of an ordinary untwisted string at $ c=26 $! This {\it spectral 
equivalence} was 
discussed in detail  for the $ \mathbb{Z}_{2} $ orbifold-string theory 
in Ref. [30], using the (inverse of the) order-two orbifold-induction procedure 
[10] To see 
the spectral equivalence here for all prime $ \lambda $ one need only 
generalize the discussion of Ref. [30], using now the 
(inverse of the) 
order-$ \lambda $ {\it orbifold induction 
procedure} [10]. As seen explicitly in the following set of equations, the general 
procedure maps the twisted matter of the extended 
Virasoro generators (2.2) and the extended physical state 
condition (2.17) to the (unhatted) equivalent conventional 
untwisted system at $ c=26 $: 
\begin{subequations} 
\begin{gather} 
  L(\lambda m + r) \equiv \lambda {\hat L}_r \left( {m + \tfrac{r}{\lambda }} \right) - \frac{{13}}
{{12}}(\lambda ^2  - 1)\delta _{m + \frac{r}
{\lambda },0}  \hfill \\
  J_\mu  (\lambda m + r) \equiv J_{r\mu } \left( {m + \tfrac{r}
{\lambda }} \right) \hfill \\ 
  L(M) =  - \frac{{\eta ^{\mu \nu } }}
{2}\sum_{P \in \mathbb{Z}} {\normalorder J_\mu  (P)J_\nu  (M - P)\normalorder_M ,\quad M \in \mathbb{Z}}  \hfill \\ 
  [J_\mu  (M),J_\nu  (N)] =  - \eta _{\mu \nu } M\delta _{M + N,0}  \hfill \\ 
  [L(M),J_{\mu}(N)] = -N J_{\mu}(M+N) \hfill \\ 
  [L(M), L(N)] = (M-N) L(M+N) + \tfrac{26}{12}M(M^{2}-1)\delta_{M+N, 
  0} \hfill \\
  (L(M \geq 0) - \delta _{M,0} )\left| \chi  \right\rangle  = 0. 
\end{gather}
\end{subequations}
A right-mover copy of this system is similarly obtained , and the 
conventional physical 
states $\{|\chi\rangle\}$ in Eq. (2.18g)  are exactly the 
{\it same} physical states, with the same physical spectrum, as those 
defined by the extended physical state condition 
(2.17) -- each state $|\chi\rangle $ having now been reexpressed via Eq. (2.18b) in terms 
of the untwisted modes $\{ J,\,\bar{J}\} $. As emphasized for the case 
$ \lambda=2$ in Ref. [27], this spectral equivalence holds  
for the $ \hat{c}=26\lambda $ and $ c=26 $ systems {\it only} when they are considered as string theories 
(truncated respectively by the  physical state conditions (2.17) 
and/or 
(2.18g)), and {\it not} for the CFT's themselves. We shall revisit this 
 equivalence from other viewpoints in the next subsection and Sec. 3.


\subsection{Example: Four-Point Amplitude in $\mbox{\boldmath 
{$U(1)^{52}/\boldsymbol{\mathbb{Z}}_2$}}$}

As an explicit example at $\lambda=2$, I will evaluate the twisted four-point string 
amplitude (2.9) for the single twisted sector of the permutation 
orbifold $(U(1)^{26}\times U(1)^{26})/\mathbb{Z}_2 $ at $\hat c=52 $. Towards 
this, we need the explicit form of the twisted vertex operators 
[19] for this case:
\begin{subequations} 
\begin{gather} 
  {\hat g}(\mathcal{T},{\bar z},z) =\,\, {\hat g}_ -  (\mathcal{T},{\bar 
  z})\,{\hat g}_ +  (\mathcal{T},z) \hfill \\
\begin{split} 
  {\hat g}_ +  (\mathcal{T},z)\, =\,\, & z^{ - \frac{1}
{2}} e^{ - i\tau _0 T_\mu  {\hat q}^\mu  } e^{ - \frac{1}
{2}\tau _0 T \cdot {\hat J}_0 (0)}  \,\times \\
   & \times \exp(\frac{1}
{2}\tau _0 T \cdot \sum\nolimits_{m \ne 0} {{\hat J}_0 (m)\frac{{z^{ - m} }}
{m}}) \, \times  \\ 
   & \times  \exp(\frac{1}
{2}\tau _1 T \cdot \sum\nolimits_m {{\hat J}_1 \left( {m + \tfrac{1}
{2}} \right)\frac{{z^{ - \left( {m + \tfrac{1}
{2}} \right)} }}
{{m + \frac{1}
{2}}}})  \hfill \\ 
\end{split}\\ 
  \left[ {{\hat q}^\mu  ,{\hat J}_{r\nu } \left( {m + \tfrac{r}
{\lambda }} \right)} \right] = i\delta _\nu ^\mu  \delta _{m + \frac{r}
{\lambda },0} ,\quad\,\, [{\hat q}^\mu  ,{\hat q}^\nu  ] = 0\,. \hfill
\end{gather}
\end{subequations}
Here I have used $ t_{0}=\tau_{0}=\thickone_{2} $ and 
$t_{1}=\tau_{1}$ (the first Pauli matrix), and dot products are defined 
with the target-space metric $\eta $ in Eq. (2.2e). The
form of the right-mover $ \hat{g}_{-} $ is the same as $\hat{g}_{+}$ 
with $z \to \bar z$, $\hat J \to \hat{\bar J}$ and $\hat q \to \hat{\bar 
q}$. Then it is straightforward to compute the orbifold correlator
\begin{multline} 
  \langle  - T^{(4)} |R({\hat g}(\mathcal{T}^{(3)} ,{\bar z}_3 ,z_3 ){\hat g}(\mathcal{T}^{(2)} ,{\bar z}_2 ,z_2 ))|T^{(1)} \rangle  \\
  = \thickone_2 \delta ^{26} \left( {\sum_{i = 1}^4 {T^{(i)} } } \right)|z_3 |^{ - \left( {T^{(3)}  \cdot T^{(2)}  + 1} \right)} |z_2 |^{ - \left( {T^{(2)}  \cdot T^{(1)}  + 1} \right)}  \times \\
   \times |z_3  - z_2 |^{ - T^{(3)}  \times T^{(2)} } \left| {\frac{{\sqrt {z_3 }  - \sqrt {z_2 } }}
{{\sqrt {z_3 }  + \sqrt {z_2 } }}} \right|^{ - T^{(3)}  \cdot T^{(2)} } 
\end{multline}
which is proportional to the unit matrix $\thickone_{2}$ because only even powers of 
$\tau_{1}$ survive the contraction.


The corresponding four-point orbifold-string amplitude 
\begin{multline} 
  {\hat A}_4^{(2)} (\{\mathcal{T}\} ) = \delta ^{26} \left( {\sum_{i = 1}^4 {T^{(i)} } } \right)S^T \thickone_2 S \times \\
   \times \int {\frac{{d^2 z}}
{{8\pi }}\,\,|z|^{ - \left( {T^{(2)}  \cdot T^{(1)}  + 1} \right)} |1 - z|^{ - 
T^{(2)}  \cdot T^{(3)} } }\times   \\
  \times \left( {\left| {\frac{{1 - \sqrt z }}
{{1 + \sqrt z }}} \right|^{ - T^{(2)}  \cdot T^{(3)} }  +\,\,\, \left| {\frac{{1 + \sqrt z }}
{{1 - \sqrt z }}} \right|^{ - T^{(2)}  \cdot T^{(3)} } } \right)
\end{multline}
is then obtained from Eq. (2.13) at $\lambda=2 $. According to Eq. (2.10), 
the spinor factor $S^{T}\thickone_2 S$ here is 
trivial -- and moreover the identities
\begin{equation}|1 - z|^a \left| {\frac{{1 \mp \sqrt z }}
{{1 \pm \sqrt z }}} \right|^a  = |1 \mp \sqrt z |^{2a} \end{equation}
can be used to further simplify this result. The  square roots in 
these equations reflect the half-integer moding of the 
extended Virasoro generators (2.2a) and twisted vertex operators (2.19) of the 
twisted sector in this case.


In fact, all the roots in the expression (2.21) can be removed by the following change of 
variable
\begin{subequations} 
\begin{gather} 
  z = u^2 ,\quad d^2 z = 4|u|^2 d^2 u \\ 
  \int_0^\pi  {d^2 u\:|u|^b (|1 - u|^a  + |1 + u|^a ) = \int_{ - \pi 
  }^\pi  {d^2 u\:|u|^b |1 - u|^a } } 
\end{gather}
\end{subequations}
where the range of $\text arg(u)$ is shown on the integrals.  This brings us 
to our final result:
\begin{subequations} 
\begin{gather}
\hat{A}_{4}^{(2)} (\{ \mathcal{T}\} ) = \hat{A}^{(2)}_4(\{T\})_I,\quad\quad\quad I=0, 1 \\ 
 \quad =\delta ^{26} \left( {\sum_{i = 1}^4 {T^{(i)} } } \right) 
 \int_{-\pi}^{\pi} {\frac{{d^2 u}} 
{{2\pi }}\,\,|u|^{ - T^{(2)}  \cdot T^{(1)} } |1 - u|^{ - 2T^{(2)}  \cdot T^{(3)} } }.
\end{gather}
\end{subequations}
If we identify the dimensionful momenta $\{k\}$ and the closed-string 
Regge slope $\alpha'_c$  as
\begin{equation}
T = \sqrt {\alpha '_c}\,k ,\quad\quad \left( {T^{(i)} } \right)^2  = \alpha '_c \left( {k^{(i)} } \right)^2  =  - 2
\end{equation}
then the integral in Eq. (2.24b) is recognized as the Virasoro-Shapiro 
amplitude [10,12] of the ordinary critical unoriented closed string.



Our interpretation of the result (2.24) is shown as the $\lambda=2$ 
case of Fig. 2 (see 
Subsec. 2.2 and Fig. 1): For each type $\{I=0,1\}$ of 
external untwisted tachyon, the cross-channel of the $I$-independent 
amplitude $\hat{A}_{4I}^{(2)}$ shows the spectrum of an ordinary 
untwisted closed 
string of type $I$. This duality is perhaps not surprising since we 
already knew that a) the external untwisted tachyons had the 
same mass as the the ground-state tachyon of the twisted sector, b) the 
physical spectrum 
of the twisted sector is identical to that of an ordinary closed 
string, and c) the twisted trees in Eq. (2.12b) are symmetrized with respect to the 
 vertex operators $\hat{g}_{I}$ .
I emphasize that the external tachyons of type $I$ couple only to the 
internal untwisted closed string of type $I$ -- which is consistent with 
the fact that (before symmetrization) the untwisted 
sector $U(1)^{52}=(U(1)^{26}\times U(1)^{26})$ of the orbifold is nothing but 
two {\it decoupled} copies of the untwisted closed string.
 
\begin{figure}[htbp]
	\centering
		\includegraphics[angle=0,width=0.6\columnwidth]{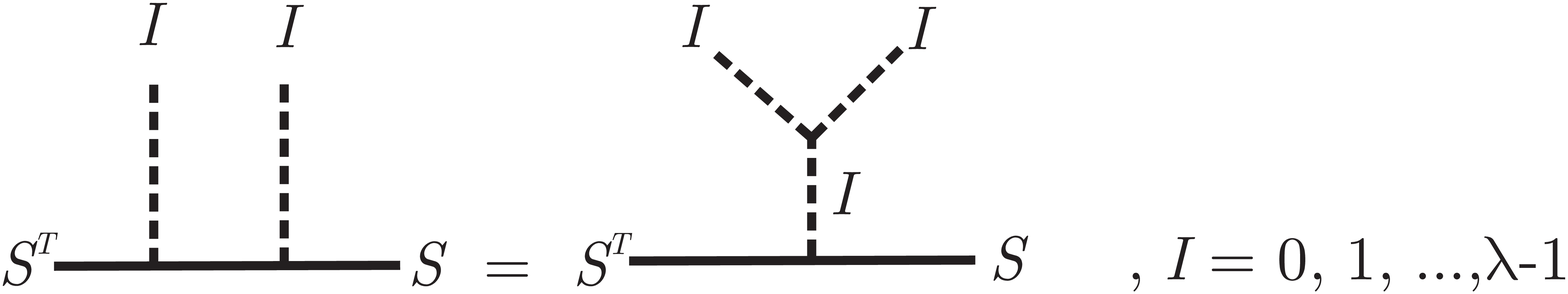}
	\caption{The cross channel is the untwisted sector.}
	\label{crosschannel}
\end{figure}

The principles of the previous paragraph hold for emission of 
an arbitrary number of untwisted tachyons of type $I$, and it is straightforward to check
that the $I$-independent twisted trees $\{\hat{A}_{nI}^{(2)}\}$ are also proportional 
to  ordinary Virasoro-Shapiro n-point amplitudes. For brevity I will 
not give these computations here but, as illustrated in our 
four-point example above, the steps are nothing but a complexified version of 
those given in the  n-point twisted open-string computation of 
Ref. [31]. Given the known decoupling of zero-norm states in the 
Virasoro-Shapiro trees, this provides the expected [28] no-ghost 
theorem for the twisted sector of the $ \mathbb{Z}_2 $ 
permutation-orbifold string.  Indeed, except that $I=0,\ldots,\lambda-1$, the principles of the 
previous paragraph hold for the orbifolds with all prime $\lambda$, so 
I expect the {\it same} Virasoro-Shapiro amplitudes 
$\hat{A}_{nI}^{(\lambda)}\sim \hat{A}_{nI}^{(2)}$ for $\lambda \geq 3$. 
This expectation should be checked directly from the twisted trees 
(2.9),  using the explicit forms of the (higher-fractional-moded) vertex 
operators given in Ref. [19].

\newpage 
\section{The Twisted Sectors on the Torus}
\subsection{The Cosmological Constant $\mbox{\boldmath{$\Lambda$}}$ }

To establish a language for the development below, I begin with a 
short {\em review} of the cosmological constant (see e.g. Ref. [12]).

In $D$-dimensional Euclidean space, the {\it naive} or {\it particle-theoretic}
 form of the one-loop cosmological constant (ground state 
energy/volume) for any relativistic theory is 

\begin{subequations} 
\begin{gather} 
  \Lambda _N  = \frac{{E_0 }}
{V} = \sum_j {\int {(d^{D - 1} k)} \,\,\frac{1}
{2}\omega _j \left( {\vec k} \right)}  =  - \frac{1}
{2}\int_0^\infty  {\frac{{dt}}
{t}\int {(d^D k)\sum_j {e^{ - t(k^2  + m_j^2 )} } } }  \hfill \\ 
  (d^D k) = \frac{{d^D k}}
{{(2\pi )^D }},\quad \omega _j \left( {\vec k} \right) = \sqrt {\vec k^2 + m_j^2 } ,\quad k^2  = k_0^2  + \vec k^2 
\end{gather}
\end{subequations} 
where the sum is over all {\it particles} $\{j\}$ in the theory. For the 
ordinary untwisted closed string in $D=26$ Euclidean dimensions, we recall that
\begin{subequations}\begin{gather} 
  L(0) = \frac{{J^2 (0)}}{2} + R,
\quad {\bar L}(0) = \frac{J^2 (0)}{2} + {\bar R},
\quad J(0) = \sqrt {\alpha '_c }\, k \hfill \\ 
  D = \frac{{\delta _{L(0),{\bar L}(0)} }}
{{L(0) + {\bar L}(0) - 2}},\quad \delta _{L(0),{\bar L}(0)}  = \int_{ - \pi }^\pi  {\frac{{d\phi }}
{{2\pi }}e^{i\phi (L(0) - {\bar L}(0))} }  
\end{gather}
\end{subequations}
where operator $D$ is the string propagator. Then Eq. (3.1) gives the 
naive or particle-theoretic form of the string cosmological constant


\begin{subequations} 
\begin{gather}
\begin{split}
  (\alpha '_c )^{13} \Lambda _N  & =  - \frac{1}
{2}\int_0^\infty  {\frac{{dt}}
{t}\int {(d^{26} J(0))}\, Tr} \left( {e^{ - t(L(0) + {\bar L}(0) - 2)} \delta _{L(0),{\bar L}(0)} } \right) \\
&   =  - \frac{1}
{2}\int_{F_N } {\frac{{d^2 \tau }}
{{(\operatorname{Im} \tau )^2 }}\,Z({\bar \tau },\tau ),\quad \tau  \equiv \frac{1}
{{2\pi }}(\phi  + it)}
\end{split} \\ 
\begin{split}
  Z(\tau ,{\bar \tau} ) & = \operatorname{Im} \tau \int {(d^{26} 
  J(0))\,Tr\left( {e^{2\pi i\tau (L(0) - 1)} e^{ - 2\pi i{\bar \tau }({\bar L}(0) - 1)} } \right)} \\
  & = \frac{1}
{{(8\pi ^2 )^{13} }}\frac{1}
{{(\operatorname{Im} \tau )^{12} }}\,|\eta (e^{2\pi i\tau } )|^{ - 48}
\end{split}\\ 
F_N :\quad|\operatorname{Re} \tau | \leq \frac{1}
{2},\quad \operatorname{Im} \tau  \geq 0 
\end{gather}\end{subequations}
which is integrated over the {\it naive} or {\it particle-theoretic} region $F_{N}$ 
of moduli space. Here $ Z(\tau ,{\bar \tau} ) $ is the modular-invariant $c=26$
 closed-string partition function, $\eta (e^{2\pi 
i\tau })$ is the Dedekind $\eta$-function, and I have assumed that 
the traces in Eq. (3.3) sum only over the physical states. The 
summation over particle-theoretic contributions described here is an 
example of the naive, operator sewing [34,12] of any string loop, and the 
same naive region $F_{N}$ is encountered as well in the naive sewing of 
closed-string loops with insertions.


Dividing by $SL(2,\mathbb{Z})$, we then obtain the {\it physical} or 
{\it string-theoretic} cosmological constant $\Lambda$
\begin{subequations} 
\begin{gather}
  F_N  \to \quad F:|\operatorname{Re} \tau | \leq \frac{1}
{2},\quad \operatorname{Im} \tau  \geq 0,\quad |\tau | \geq 1 \hfill \\ 
  (\alpha '_c )^{13} \Lambda _N \to \quad (\alpha '_c )^{13} \Lambda  =  - \frac{1}
{2}\int_F {\frac{{d^2 \tau }}
{{(\operatorname{Im} \tau )^2 }}}\, Z(\tau ,{\bar \tau }) 
\end{gather}
\end{subequations}
of the untwisted closed string, where $F$ is the standard fundamental region. If so desired, we may formally 
return to the naive, particle-theoretic form ($F\rightarrow F_{N}, \Lambda\rightarrow 
\Lambda_{N})$ by ignoring the last (circular arc) constraint in Eq. 
(3.4a) -- and indeed this return is necessary to see the particle content 
(3.1) of the theory.


\subsection{The Twisted Contribution to $ \mbox{\boldmath{$\hat \Lambda$}}$ }

In this subsection I compute the contribution of 
the twisted sectors of the permutation orbifold $U(1)^{26 \lambda}/\mathbb{Z}_\lambda$ to the 
orbifold cosmological constant $\hat \Lambda$. 

For each twisted sector, we will need the following identities:
\begin{subequations} 
\begin{gather}
{\hat L}_0 (0) = \tfrac{{{\hat J}^2 (0)}}
{{2\lambda }} + {\hat R},\quad {\hat {\bar L}}_0 (0) = 
\tfrac{{{\hat{\bar J}}^2 (0)}}
{{2\lambda }} + {\hat {\bar R}}\\ 
\delta_{{\hat L}_0 (0),{\hat {\bar L}}_0 (0)}  = \delta _{{\hat L}_0 (0),{\hat {\bar L}}_0 (0)\bmod 1} \int_{ - \pi }^\pi  {\frac{{d\phi }}
{{2\pi }}e^{i\phi ({\hat L}_0 (0) - {\hat {\bar L}}_0 (0))} } \\ 
\delta _{{\hat L}_0 (0),{\hat {\bar L}}_0 (0)\bmod 1}  = \frac{1}
{\lambda }\sum_{r = 0}^{\lambda  - 1} {e^{2\pi ir({\hat L}_0 (0) - {\hat {\bar L}}_0 (0))} }.  
\end{gather}
\end{subequations}
Recall that the ordinary Kronecker delta on the left of Eq. (3.5b) is 
also the numerator of 
the twisted propagator $\hat D$ in Eq. (2.5), and indeed the 
identities (3.5b,c) were used to obtain the integral representations of 
$\hat D$ in Eq. (2.6).


Then Eq. (3.1) gives the naive or particle-theoretic contribution $\hat{\hat 
\Lambda}_N$ of all $\lambda-1$ twisted sectors to the orbifold 
cosmological constant
\begin{subequations} 
\begin{gather}
\begin{split}
(\alpha '_c )^{13} {\hat {\hat \Lambda }}_N^{(\lambda )}  &=  - \frac{{(\lambda  - 1)}}
{2}\int_0^\infty  {\frac{{dt}}
{t}\int {(d^{26} {\hat J}(0))Tr\left( {e^{ - t({\hat L}_0 (0) + {\hat {\bar L}}_0 (0) - 2{\hat a}_\lambda  )} \delta _{{\hat L}_0 (0),{\hat {\bar L}}_0 (0)} } \right)} } \\
 &=  - \frac{{(\lambda  - 1)}}
{2}\int_{F_N } {\frac{{d^2 \tau }}
{{(\operatorname{Im} \tau )^2 }}\,\,{\hat Z}_\lambda  (\tau ,{\bar \tau })} 
\end{split}\\ 
{\hat Z}_\lambda  (\tau ,{\bar \tau }) \equiv \operatorname{Im} \tau \int {(d^{26} {\hat J}(0))Tr\left( {e^{2\pi i\tau ({\hat L}_0 (0) - {\hat a}_\lambda  )} e^{ - 2\pi i{\bar \tau }({\hat {\bar L}}_0 (0) - {\hat a}_\lambda  )} \delta _{{\hat L}_0 (0),{\hat {\bar L}}_0 (0)\bmod 1} } \right)} 
\end{gather}
\end{subequations}
where $\hat Z_{\lambda}(\tau,\bar\tau)$ will be called the {\it twisted 
partition function}. I will 
also define a physical or string-theoretic version $\hat{\hat\Lambda}$ of these 
contributions by the formal substitution $F_{N}\rightarrow F$ 


\begin{equation} 
(\alpha '_c )^{13} {\hat {\hat \Lambda }}^{(\lambda )}  \equiv  - \frac{{(\lambda  - 1)}}
{2}\int_F {\frac{{d^2 \tau }}
{{(\operatorname{Im} \tau )^2 }}\,\,{\hat Z}_\lambda  (\tau ,{\bar \tau })} 
\end{equation}
although strictly speaking there is no need to divide by  
$SL(2,\mathbb{Z})$ until the modular-invariant completion of $\hat 
Z_{\lambda}(\tau,\bar\tau)$ in Sec. 4.


The twisted partition function can be simplified by the order-$\lambda$ 
orbifold induction procedure in Eq.(2.18), which implies the following 
relations:
\begin{subequations} 
\begin{gather}
{\hat L}_0 (0) - {\hat a}_\lambda   = \frac{1}
{\lambda }(L_0 (0) - 1),\quad {\hat {\bar L}}_0 (0) - {\hat a}_\lambda   = \frac{1}
{\lambda }({\bar L}_0 (0) - 1),\quad {\hat J}(0) = J(0)\\ 
\delta _{{\hat L}_0 (0),{\hat {\bar L}}_0 (0)\bmod 1}  = \frac{1}
{\lambda }\sum_{r = 0}^{\lambda  - 1} {e^{2\pi i\frac{r}
{\lambda }(L_0 (0) - {\bar L}_0 (0))} }  = \delta _{L_0 (0),{\bar L}_0 (0)\bmod \lambda } \\ 
{\hat D} = \frac{{\delta _{{\hat L}_0 (0),{\hat {\bar L}}_0 (0)} }}
{{\lambda ({\hat L}_0 (0) + {\hat {\bar L}}_0 (0) - 2{\hat a}_\lambda  )}} = \frac{{\delta _{L_0 (0),{\bar L}_0 (0)} }}
{{L_0 (0) + {\bar L}_0 (0) - 2}} = D. 
\end{gather}\end{subequations}
I have included here the closely-related propagator identity (3.8c), 
which was in fact used to fix the scale of the twisted propagator 
$\hat D$ in Subsec. (2.2). Then we find from Eqs. (3.6b) and (3.8) that


\begin{equation} 
\begin{split}
{\hat Z}_\lambda  (\tau ,{\bar \tau }) & = \operatorname{Im} \tau \int {(d^{26} J(0))Tr\left( {e^{2\pi i\frac{\tau }
{\lambda }(L(0) - 1)} e^{ - 2\pi i\frac{{{\bar \tau }}}
{\lambda }({\bar L}(0) - 1)} \delta _{L(0),{\bar L}(0)\bmod \lambda} } \right)} \\
 & = \operatorname{Im} \left( {\frac{\tau }
{\lambda }} \right)\sum_{r = 0}^{\lambda  - 1} {\int {(d^{26} J(0))Tr\left( {e^{2\pi i\frac{{\tau  + r}}
{\lambda }(L(0) - 1)} e^{ - 2\pi i\frac{{{\bar \tau } + r}}
{\lambda }({\bar L}(0) - 1)} } \right)} } \\
 & = \sum_{r = 0}^{\lambda  - 1} {Z\left( {\tfrac{{\tau  + r}}
{\lambda },\tfrac{{{\bar \tau } + r}}
{\lambda }} \right)} 
\end{split}
\end{equation}
where  $Z(\tau,\bar\tau)$ is the $c=26$ string partition function in 
Eq. (3.3b). In passing from Eq. (3.6b) to Eq. (3.9) I have used the fact that 
the physical Hilbert spaces of the $\hat c = 26\lambda$ and the 
$c=26$ descriptions are isomorphic (see Refs. [13] and [30]) -- and 
hence the traces are identical. Up to a numerical factor of $\lambda$ 
(associated to the ``extra'' prefactor $\operatorname{Im} \tau$ in Eqs. (3.3b) and (3.6b)), 
the last line of the result (3.9) is the standard form [8] of the twisted partition 
function for the $\hat c= 26\lambda$ orbifold conformal field theories.


\subsection{Spectral Equivalence in the Cosmological Constant}


We saw in Eq. (3.9) that the partition function 
${\hat Z}_{\lambda}$ of each twisted $\hat c=26\lambda$ string is proportional to 
the standard twisted partition function of the corresponding 
permutation-orbifold 
CFT. This is natural enough, except that the spectra of the
permutation-orbifold CFT's are notoriously intricate [7,10] -- while we 
know from Ref. [30]
 and Sec. 2 that, {\em as a string 
theory} restricted by the extended physical state condition (2.17), 
the physical spectrum of each twisted sector is equivalent to that of an 
ordinary untwisted $c=26$ closed string. The question is then how to 
see this spectral equivalence at the level of the  orbifold 
cosmological constant, and the answer must lie in the final 
$\tau$-integration of Eqs. (3.6) and (3.7). Indeed the following 
results on the torus stand in  close analogy
 with those given above (see Subsecs. 2.3 and 2.4) for the 
twisted trees -- where the spectral equivalence is seen only after the 
$z$-integrations over the CFT correlators on the sphere. 

Spectral equivalence on the torus can be understood in terms of two lemmas, 
whose proof will be sketched after the statement of the lemmas. The first lemma is a formal 
integration identity 
 \begin{subequations} 
\begin{gather}
\int_{F_N } {\frac{{d^2 \tau }}
{{(\operatorname{Im} \tau )^2 }}\sum_{r = 0}^{\lambda  - 1} {Z\left( {\tfrac{{\tau  + r}}
{\lambda },\tfrac{{{\bar \tau } + r}}
{\lambda }} \right) = \int_{F_N } {\frac{{d^2 \tau }}
{{(\operatorname{Im} \tau )^2 }}\,Z(\tau ,{\bar \tau }} )} } \\ 
 \to \quad {\hat {\hat \Lambda }}_N^{(\lambda )}  = (\lambda  - 1)\Lambda _N  
\end{gather}
\end{subequations}
where $F_{N}$ is the naive region in Eq. (3.3c).
This identity tells us that, as required by spectral equivalence on the sphere, the 
integrated particle-theoretic contribution of each twisted  
$\hat c=26\lambda$ sector is the same as the integrated 
particle-theoretic contribution $\Lambda_{N}$ (see Eq. (3.31)) of an ordinary closed 
string. 


The second lemma is  a physical or string-theoretic version of the same
spectral equivalence
\begin{subequations} 
\begin{gather}
\int_{F } {\frac{{d^2 \tau }}
{{(\operatorname{Im} \tau )^2 }}\sum_{r = 0}^{\lambda  - 1} {Z\left( {\tfrac{{\tau  + r}}
{\lambda },\tfrac{{{\bar \tau } + r}}
{\lambda }} \right) = \int_{\hat F_\lambda } {\frac{{d^2 \tau }}
{{(\operatorname{Im} \tau )^2 }}\, Z(\tau ,{\bar \tau }} )} } \\ 
 \to \quad {\hat {\hat \Lambda }}^{(\lambda )}  =  - \frac{{(\lambda  - 1)}}
{2}\int_{{\hat F}_\lambda  } {\frac{{d^2 \tau }}
{{(\operatorname{Im} \tau )^2 }}\,Z(\tau ,{\bar \tau })} \,\, \ne (\lambda  - 1)\Lambda \\ 
{\hat F}_\lambda  :|\operatorname{Re} \tau | \leq \frac{1}
{2},\quad \operatorname{Im} \tau  \geq 0,\quad |\lambda \tau  + r| \geq 1,\quad r = 0, \pm 1, \ldots , \pm \left\lfloor {\tfrac{\lambda }{2}} \right\rfloor 
\end{gather}
\end{subequations}
where $F$  is the fundamental region and $\left\lfloor  x \right\rfloor$ is the floor of $x$.
This remarkable relation reconfirms at the physical level that the spectrum 
of each twisted sector of $U(1)^{26\lambda}/\mathbb{Z}_\lambda$ is that of an ordinary string, but in this 
description the contribution of each sector must be cut off by the {\it new modular region} $\hat F_\lambda$.
 
For reference, Fig. 3 shows the first few modular regions $\{\hat 
F_\lambda\}$ as the areas above the circular arcs.


\begin{figure}[htbp]
	\centering
		\includegraphics[angle=0,width=0.6\columnwidth]{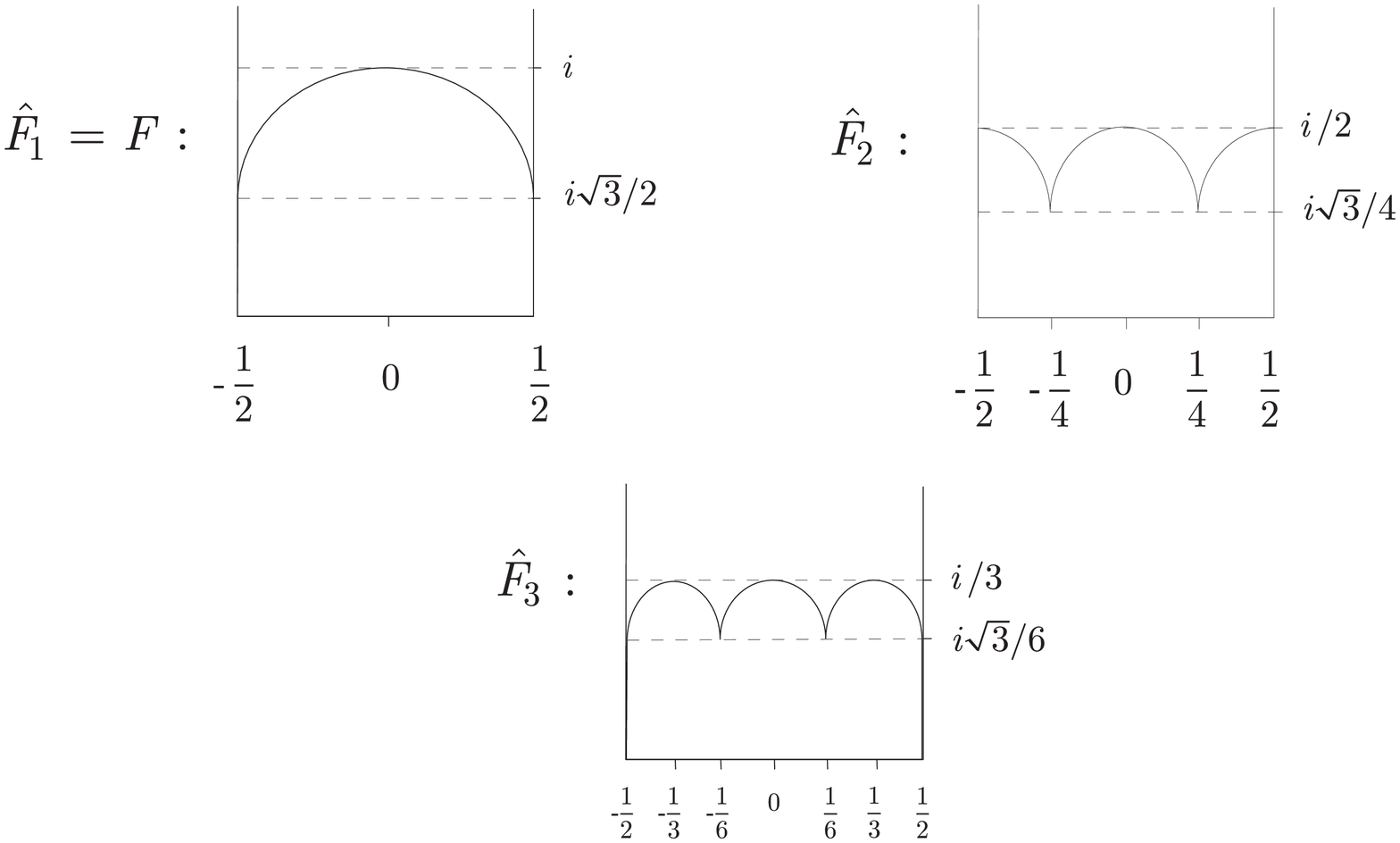}
	\caption{The first three regions $\hat F_\lambda$}
	\label{fig:regionsF}
\end{figure}


In fact, both of the lemmas in Eqs. (3.10a) and (3.11a) are nothing but ``cut and paste'' identities
which hold for any function $Z(\tau, \bar\tau)$ which is periodic with period 
one:
\begin{subequations} 
\begin{gather}
Z(\tau ' + 1,{\bar \tau }' + 1) = Z(\tau ',{\bar \tau }'),\quad \tau ' = \frac{\tau }
{\lambda }\\ 
 \to \quad {\hat Z}_\lambda  (\tau  + 1,{\bar \tau } + 1) = {\hat Z}_\lambda  (\tau ,{\bar \tau }). 
\end{gather}
\end{subequations}
This includes any modular-invariant $Z(\tau, \bar\tau)$, such as the 
partition function (3.3b) of the ordinary closed string. The proofs 
of these identities are straightforward but can 
appear somewhat involved at the algebraic level, as seen for example in the 
following sketch for the case $\lambda=2$ of the identity in Eq. (3.11a). 
Begin with the integration over the two terms

\begin{equation} 
{\hat Z}_2 (\tau ,{\bar \tau }) = Z\left( {\tfrac{\tau }
{2},\tfrac{{{\bar \tau }}}
{2}} \right) + Z\left( {\tfrac{{\tau  + 1}}
{2},\tfrac{{{\bar \tau } + 1}}
{2}} \right)
\end{equation}
on the left side of the identity, and follow the changes of variable:

\begin{subequations} 
\begin{align}
  {\text{a) }}&\tau ' = \left\{ {\begin{array}{*{20}c}
   {\tau  + 1} & , & { - \frac{\pi }
{2} < \operatorname{Re} \tau  < 0}  \\ 
   \tau  & , & {0 < \operatorname{Re} \tau  < \frac{\pi }
{2}}  \\ 
 \end{array} } \right. (\text{periodicity of }\hat Z_2) \hfill \\ 
  {\text{b) }}&\tau '' = \frac{{\tau '}}
{2}\quad {\text{(rescale)}} \hfill \\
  {\text{c) }}&\tau ''' = \tau '' - \frac{1}
{2} \quad ({\text{second term of (}}3.13){\text{ only)}}. 
\end{align}\end{subequations}
After step (a) for example, the integration region is $0\leq {\rm Re}\tau' 
\leq 1$ with a cusp at $\tfrac{1}{2}$. Both integrands are equal to 
$Z(\tau,\bar\tau)$ after the shift (c), and the integration region 
pastes to the modular region $\hat F_2$.

The proof of the particle-theoretic counterpart in Eq. (3.10a) follows 
the same steps, now ignoring the circular 
arcs of the modular regions $\{\hat F_\lambda\}$. In this case of course, the singularity of 
the integrands at $\operatorname{Im}\tau=0$ means that the steps and 
therefore the result are formal.

\newpage 

\section{About the Untwisted Contribution to $ \mbox{\boldmath{$\hat \Lambda$}}$ }

The contribution of the untwisted, 
permutation-symmetric sector of the orbifold-string system is surprisingly more difficult to understand.

\subsection{Trouble with the Standard CFT Prescription}

Given $\lambda$ copies of any conformal field theory with discrete 
spectrum and partition function $Z(\tau,\bar\tau)$, the standard CFT prescription
for the full modular-invariant partition function of the cyclic 
permutation orbifold was given
in Ref. [8].  After integration over the fundamental 
region, this prescription gives the following model 
${\hat\Lambda}_{G}$ of the full orbifold cosmological constant:
\begin{multline} 
(\alpha '_c )^{13} {\hat \Lambda }_G  =  - \frac{\lambda }
{2}\int_F \frac{{d^2 \tau }}
{{(\operatorname{Im} \tau )^2 }}\Big\{ \frac{{Z^\lambda  (\tau ,{\bar \tau })}}
{\lambda } \,+  \\ +  \frac{{\lambda  - 1}}
{\lambda }\left[ {Z(\lambda \tau ,\lambda {\bar \tau }) + \sum_{r = 0}^{\lambda  - 1} {Z\left( {\tfrac{{\tau  + r}}
{\lambda },\tfrac{{{\bar \tau } + r}}
{\lambda }} \right)} } \right] \Big\} 
\end{multline}
From our point of view, the overall scale of this model is set by its 
final sum of terms, which correctly integrates to the twisted contribution  
$\hat{\hat\Lambda}$ in Eqs. (3.7) and (3.11). The other two terms 
in Eq. (4.1) are contributions from the untwisted sector. In particular, 
the second term provides the modular-invariant completion of the twisted 
contribution, the sum of all terms in the square brackets being a 
Hecke transform of order $\lambda$. From the point of view of Ref. 
[8], the first term is the original unsymmetrized partition function $\prod_I 
Z_I = Z^\lambda$ of the untwisted sector, and the second term is a 
correction to the first term for the symmetrized counting -- using the discreteness of 
the spectrum -- of the so-called ``diagonal'' states
  $\{|x\rangle \otimes \ldots \otimes |x\rangle\}$  
 of the untwisted sector.  
 In our discussion above  we have 
 assumed for simplicity that the spectrum of the ordinary closed-string 
 partition function $Z(\tau,\bar\tau)$
 is not discrete (hence the extra powers of $\operatorname{Im} 
\tau$ in Eq. (3.3b)), but this mismatch can be remedied by compactification.


There is of course no objection to the partition function of Ref. 
[8] as an orbifold CFT, and indeed I had expected the form (4.1) to provide
 the desired full cosmological constant 
of the orbifold --
 but in fact there is a physical problem  
with the untwisted contribution to this model {\it as a string system}.
 The point is that the ordinary closed-string
 partition function $Z(\tau,\bar\tau)$ in Eq. (3.3b) involves a single momentum integration
\begin{equation} 
	Z = \int {(d^{26} J(0)) \ldots } \quad  = \int {(d^{26} k) \ldots } 
\end{equation}
and this gives an untwisted contribution to the model   
which is proportional to {\it $\lambda$ momentum integrations} because
\begin{equation} 
Z^\lambda   = \int {\left( {\prod\limits_{I = 0}^{\lambda  - 1} {(d^{26} 
k_I )} } \right) \ldots\,\,\,. } 
\end{equation}
But this {\it higher-loop term} in ${\hat\Lambda}_{G}$ contradicts the standard one-loop 
particle-theoretic form (3.1) of any cosmological constant!

On the basis of this simple observation, we are forced to conclude  
for all prime $\lambda \geq 1$ that 
the untwisted contribution (and hence all the contributions) to the 
orbifold cosmological constants of  $U(1)^{26 \lambda}/\mathbb{Z}_\lambda$  must be
{\it linear} in the  partition function $Z(\tau,\bar\tau)$ of the ordinary closed string.  
Although the constructions of this paper will be complete only for 
these particular  
orbifolds, I emphasize that the 
same conclusion (linear in $\int(d^{26}k))$ applies as well to the cosmological constant of
the general permutation orbifold in Eq. (1.1).


\subsection{Particle-Theoretic Form of\\
\hspace*{.4in} the Untwisted Contribution}

We can further support the conclusion of the previous subsection by 
considering the untwisted contribution to the cosmological constant
in an analogous set of permutation-invariant quantum field theories. 
Consider in particular the $\mathbb{Z}_\lambda\text{-invariant}$ Hamiltonian
\begin{equation} 
	H = \int {d^{25} x\left[ {\sum_j {\sum_{r = 0}^{\lambda  - 1} {\frac{1}
{2}\left( {\pi _{jr}^2  + |\nabla \phi _{jr} |^2  + m_j^2 \phi _{jr}^2 } 
\right)} } } \right]\,\,
 +\quad V_\lambda  [\phi ]} 
\end{equation}
where $V_{\lambda}[\phi]$ is any interaction which is invariant under 
cyclic permutations  $\{\phi_{jr}\rightarrow \phi_{j,r+1}\}$ of the 
copies $\{r\}$
of each particle $j$. At each value of $\lambda$, the one-loop contribution 
$\Lambda_{\lambda}$ to the exact cosmological constant
\begin{subequations} 
\begin{gather}
	\Lambda _{{\text{exact}}}  = \,
	\langle 0 | (H /V) | 0 \rangle\, = \,\Lambda _\lambda   + ({\text{higher loops}})\\
\Lambda _\lambda   = \sum_j {\sum_{r = 0}^{\lambda  - 1} {\int {(d^{25} k)\frac{1}
{2}\sqrt {\vec k^2  + m_j^2 } } } } 
\end{gather}\end{subequations}
is independent of the interaction and {\it automatically symmetric} under 
$\mathbb{Z}_{\lambda}$. Then we find that


\begin{subequations} 
\begin{gather}
\Lambda _\lambda   = \lambda \Lambda \\  
\Lambda  =  - \frac{1}
{2}\int_0^\infty  {\frac{{dt}}
{t}\int {(d^{26} k)\sum_j {e^{ - t(k^2  + m_j^2 )} } } } 
\end{gather}
\end{subequations}
where $\Lambda$ is the one-loop cosmological constant for a single 
copy of all the particles $\{j\}$ in the theory.

Promoting Eq. (4.6) to the string theory (see Subsec. 2.1) of 
$\lambda$ copies of $U(1)^{26}$,
we learn that the untwisted  particle-theoretic contribution 
to the orbifold cosmological constant is indeed {\it linear} in the ordinary 
closed-string 
partition function
\begin{equation}  
	(\alpha '_c )^{13} {\hat \Lambda }_N^{({\text{sym}})}  =  - \frac{\lambda }
{2}\int_{F_N } {\frac{{d^2 \tau }}
{{(\operatorname{Im} \tau )^2 }}\,Z(\tau ,{\bar \tau })} 
\end{equation}
where $Z(\tau,\bar\tau)$ and the naive region $F_{N}$  are given Eq. (3.3). At 
this level then, the one-loop contribution ${\hat \Lambda }_N^{({\text{sym}})}=\lambda \Lambda_{N}$ of the
 untwisted sector is equivalent to that of
 $ \lambda $ ordinary closed strings, and this is consistent with the cross-channel
behaviour of the twisted trees (see Secs. 2 and 6). The reasoning 
above is easily extended to see that ${\hat \Lambda 
}_N^{({\text{sym}})}=K\Lambda_{N}$ 
for the untwisted particle-theoretic contribution to the general permutation orbifold 
$U(1)^{26K}/H(\text{perm})$ on $K$ copies.


 The particle-theoretic result in Eq. (4.7) is in fact a natural structure in the
 permutation-orbifold conformal field theory [8,13]. In 
particular, the following formal integration identity 
\begin{equation} 
	\lambda \int_{F_N } {\frac{{d^2 \tau }}
{{(\operatorname{Im} \tau )^2 }}}\, Z(\tau ,{\bar \tau }) = \int_{F_N } {\frac{{d^2 \tau }}
{{(\operatorname{Im} \tau )^2 }}\,Z(\lambda \tau ,\lambda {\bar \tau }} ).
\end{equation}
shows that, at this naive level, summing over $\lambda$ ordinary 
strings is equivalent to summing over the so-called {\it diagonal 
states} 
$\{|x\rangle \otimes \ldots \otimes |x\rangle\}$
of the untwisted sector of the orbifold. The identity (4.8) provides 
yet another {\it spectral 
equivalence}, which parallels the role of Eq. (3.10a) in the twisted sector. 
The identity itself can be proven directly from the explicit form of 
$Z(\tau,\bar\tau)$ in Eq. (3.3b), or by  the following formal steps

\begin{subequations} 
\begin{gather}
\int_{ - 1/2}^{1/2} {d(\operatorname{Re} \tau )e^{2\pi i\lambda \operatorname{Re} \tau (L(0) - {\bar L}(0))} }  = \delta _{L(0),{\bar L}(0)} \\ 
\int_{F_N } {\frac{{d^2 \tau }}
{{(\operatorname{Im} \tau )^2 }}Z(\lambda \tau ,\lambda {\bar \tau }) = \int_0^\infty  {\frac{{d(\operatorname{Im} \tau )}}
{{(\operatorname{Im} \tau )^2 }}f(\lambda \operatorname{Im} \tau ) = \lambda \int_0^\infty  {\frac{{d(\operatorname{Im} \tau )}}
{{(\operatorname{Im} \tau )^2 }}f(\operatorname{Im} \tau )} } } \\ 
f(\operatorname{Im} \tau ) \equiv\, \operatorname{Im} \tau \int {(d^{26} J(0))Tr\left( {e^{ - 2\pi \operatorname{Im} \tau (L(0) + {\bar L}(0) - 2)} \delta _{L(0),{\bar L}(0)} } \right)} 
\end{gather}
\end{subequations}
from the trace form of $Z(\tau,\bar\tau)$ in the same equation. It remains to 
reconcile these conclusions with modular invariance in the full 
orbifold cosmological constant $\hat\Lambda$.

\newpage 

\section{Provisional Forms of\\ 
\hspace*{.4in}the Cosmological Constant}

\subsection {Derivation from the Spectral Equivalences}

In this section, I will consider the following 
{\it one-parameter $\beta$-family} of candidates for the {\it full} orbifold cosmological 
constant
\begin{multline} 
(\alpha '_c )^{13} {\hat\Lambda ^{(\lambda )}} (\beta )  =  - \frac{1}
{2}\int_F \frac{{d^2 \tau }}
{{(\operatorname{Im} \tau )^2 }}\Big{\{} (2\lambda  - 1 - \beta (\lambda  + 
1))Z(\tau ,{\bar \tau }) \,+ \\
+ \beta \left[ {Z(\lambda \tau ,\lambda {\bar \tau }) + \sum_{r = 0}^{\lambda  - 1} {Z\left( {\tfrac{{\tau  + r}}
{\lambda },\tfrac{{{\bar \tau } + r}}
{\lambda }} \right)} } \right]  \Big{\}}.
\end{multline}
of the permutation-orbifold-string system $ U(1)^{26\lambda}/\mathbb{Z}_\lambda $ at $\hat c=26\lambda$.
The fundamental region $F$ of moduli space is defined in Eq. (3.4).
Note that the integrand of this expression is modular-invariant for each prime 
$\lambda $ at any value of the arbitrary complex parameter $\beta$. 
Moreover, the integrand is independent of $\beta$ when $\lambda=1$, so 
that the provisional form (5.1) includes the ordinary cosmological 
constant $\Lambda = {\hat\Lambda}^{(1)}$ (see Eq. (3.4b)) of the  
untwisted closed string.  It will 
be instructive to begin with the formal derivation of this family from the particle-theoretic 
contributions, returning in the following subsection to discuss special choices of $\beta$.


Collecting the twisted and untwisted particle-theoretic contributions 
in Eqs. (3.6a), (3.9) and (4.7), we find the naive, operator-sewn form of the full orbifold cosmological 
constant
\begin{equation} 
{\hat \Lambda }_N^{(\lambda )}  =  - \frac{1}
{2}\int_{F_N } {\frac{{d^2 \tau }}
{{(\operatorname{Im} \tau )^2 }}\left\{ {\lambda Z(\tau ,{\bar \tau }) + (\lambda  - 1)\sum_{r = 0}^{\lambda  - 1} {Z\left( {\tfrac{{\tau  + r}}
{\lambda },\tfrac{{{\bar \tau } + r}}
{\lambda }} \right)} } \right\}} 
\end{equation}
where $F_{N}$ is the naive integration region in Eq. (3.3c). In this form the 
integrand is not modular invariant, but in fact we can change the integrand 
by the formal integration identities
\begin{subequations} 
\begin{gather}
\int_{F_N } {\frac{{d^2 \tau }}
{{(\operatorname{Im} \tau )^2 }}\,Z(\tau ,{\bar \tau })}  = \int_{F_N } {\frac{{d^2 \tau }}
{{(\operatorname{Im} \tau )^2 }}\left\{ {\alpha Z(\tau ,{\bar \tau }) + \frac{\beta }
{\lambda }Z(\lambda \tau ,\lambda {\bar \tau })} \right\}} \\ 
\int_{F_N } {\frac{{d^2 \tau }}
{{(\operatorname{Im} \tau )^2 }}\sum_{r = 0}^{\lambda  - 1} {Z\left( {\tfrac{{\tau  + r}}
{\lambda },\tfrac{{{\bar \tau } + r}}
{\lambda }} \right)} }  = \int_{F_N } {\frac{{d^2 \tau }}
{{(\operatorname{Im} \tau )^2 }}\left\{ {\gamma \sum_{r = 0}^{\lambda  - 1} {Z\left( {\tfrac{{\tau  + r}}
{\lambda },\tfrac{{{\bar \tau } + r}}
{\lambda }} \right)}  + \delta Z(\tau ,{\bar \tau })} \right\}} \\ 
\alpha  + \beta  = \gamma  + \delta  = 1 
\end{gather}
\end{subequations}
which follow straigntforwardly from the formal spectral-equivalence relations 
(4.8) and (3.10a). Here $\alpha,\beta,\gamma$ and $\delta$ are new complex parameters 
constrained only by Eq. (5.3c). The further requirement $\gamma=\beta/(\lambda-1)$ 
selects the following $\beta$-family of {\it modular-invariant} integrands


\begin{multline} 
(\alpha '_c )^{13} {\hat \Lambda }_N^{(\lambda )}  =  - \frac{1}
{2}\int_{F_N } \frac{{d^2 \tau }}
{{(\operatorname{Im} \tau )^2 }}\Big{\{} (2\lambda  - 1 - \beta (\lambda  + 
1))Z(\tau ,{\bar \tau })\,+ \\+ \beta \left[ {Z(\lambda \tau ,\lambda {\bar \tau }) + \sum_{r = 0}^{\lambda  - 1} {Z\left( {\tfrac{{\tau  + r}}
{\lambda },\tfrac{{{\bar \tau } + r}}
{\lambda }} \right)} } \right] \Big{\}}
\end{multline}
each of whose integrals is in fact equal to the $\beta$-independent naive form 
in Eq. (5.2). The final result (5.1) for the full physical orbifold cosmological constant then follows 
by the standard division $F_N \to F$ and $\hat \Lambda_N \to \hat \Lambda$ 
by $SL(2,\mathbb{Z})$.

This formal derivation emphasizes that the $\beta$-ambiguity in the 
physical result (5.1) is a consequence of the {\it formal spectral 
equivalences} discussed above, the entire $\beta$-family $\hat \Lambda 
(\beta)$  of provisional cosmological constants in Eq. (5.1) being rooted  $(F\to F_N)$ in 
the {\it same, $\beta$-independent sum (5.2) of particle-theoretic contributions} to the
naive cosmological constant $\hat \Lambda_N$.


One may try to push the formal steps of this derivation a bit further:
Using the spectral-equivalence identities (4.8) and (3.10a) again, one 
reconfirms
that the final form (5.4) of the naive 
cosmological constant is indeed independent of $\beta$ 
\begin{equation} 
\begin{split}
(\alpha '_c )^{13}  \frac{{\partial {\hat \Lambda }_N^{(\lambda )} }}
{{\partial \beta }} & =  \\
 =- \frac{1}
{2}&\int_{F_N } {\frac{{d^2 \tau }}
{{(\operatorname{Im} \tau )^2 }}} \left\{ { - (\lambda  + 1)Z(\tau ,{\bar \tau }) + Z(\lambda \tau ,\lambda {\bar \tau }) + \sum_{r = 0}^{\lambda  - 1} {Z\left( {\tfrac{{\tau  + r}}
{\lambda },\tfrac{{{\bar \tau } + r}}
{\lambda }} \right)} } \right\} \\
 & \hspace{-11.5mm} =0 
\end{split}
\end{equation}
and then the modular invariance of this integrand apparently implies 
$\beta$-independence of the physical cosmological constant
 $0 = \infty  \cdot (\partial {\hat \Lambda }/\partial \beta )$. This formal 
 conclusion is thought-provoking, but I have been unable to verify it 
 directly from the physical result (5.1) -- and will not assume it in the discussion below.


\subsection {Cases, Characters and So On}

The one-parameter $\beta$-family $\hat\Lambda (\beta)$ of provisional 
forms (5.1) leaves us with un embarras de richesse, while 
each member of 
the family is itself unfamiliar. Since all the members of the 
family are formally associated to the same particle-theoretic 
contributions, it will be important to 
further test the provisional forms against other string-theoretic and 
conformal-field-theoretic intuitions.  This includes in particular 
the $\beta$-dependent question
 whether any of these integrands have (or need to 
have) a {\em consistent interpretation at the level of characters} 
[7,13,35,36].

In this connection, I mention some special cases of $\hat\Lambda 
(\beta)$, starting with the case which is clearly favored by its simplicity:
\begin{subequations} 
\begin{gather}
	\beta  = 0:\quad (\alpha '_c )^{13} {\hat \Lambda }^{(\lambda )}  =  - \frac{{2\lambda  - 1}}
{2}\int_F {\frac{{d^2 \tau }}
{{(\operatorname{Im} \tau )^2 }}\,Z(\tau ,{\bar \tau })} \\ 
{\hat \Lambda }^{(\lambda )}  = (2\lambda  - 1)\Lambda. 
\end{gather}\end{subequations}
This form corresponds to the choices  $\alpha = \delta = 1$  and 
$\gamma=0$ in the derivation above, so that the formal integration 
identities in Eqs. (5.3a) and (5.3b)  reduce to the primitive spectral 
equivalences (4.8) and (3.10a). Then it is clear that Eq. (5.6) shows
the contributions of $\lambda$ ordinary closed strings from the untwisted sector and 
$\lambda-1$ ordinary closed strings from the twisted sectors, and an 
ordinary untwisted character-theoretic interpretation of this case follows after 
compactification.


At the other extreme, I note the case $\beta  = (2\lambda  - 1)/(\lambda  + 1)$
\begin{equation} 
	(\alpha '_c )^{13} {\hat \Lambda }^{(\lambda )}  =  - \frac{{2\lambda  - 1}}
{{2(\lambda  + 1)}}\int_F {\frac{{d^2 \tau }}
{{(\operatorname{Im} \tau )^2 }}\left[ {Z(\lambda \tau ,\lambda {\bar \tau }) + \sum_{r = 0}^{\lambda  - 1} {Z\left( {\tfrac{{\tau  + r}}
{\lambda },\tfrac{{{\bar \tau } + r}}
{\lambda }} \right)} } \right]} 
\end{equation}
which shows only the Hecke transform of order $\lambda$. It is 
important to note that this form is also amenable to character 
analysis after compactification : Indeed, the  existence of the 
modular-invariant Hecke transform 
tells us that even for the larger set of WZW cyclic permutation orbifold 
CFT's [13]


\begin{equation} 
	\frac{{G^{(1)}  \times  \cdots  \times G^{(\lambda )} }}
{{Z_\lambda  }},\;\lambda {\text{ prime}}\quad\quad (G \to U(1)^{26} \;{\text{for our abelian case}})
\end{equation}
there must exist a hitherto-unnoticed {\it modular-covariant 
subset} of orbifold characters which corresponds to the Hecke form.
Consulting Ref. [13], we recall that the so-called off-diagonal 
characters of the untwisted sector are responsible for the unwanted 
non-linear term $Z^{\lambda}$ of the standard partition function -- so
the desired modular-covariant ``Hecke-truncation'' $\mathcal H$ of 
the orbifold characters 
includes {\it all} the twisted characters but {\it only} 
the diagonal characters $\{\chi_{p}(\lambda\tau)\}$ of the untwisted 
sector. 


The truncated subset $\mathcal H$ can 
also be used to construct new non-abelian permutation orbifold CFT's 
beyond the standard prescription of Ref. [8]. We can be more 
explicit for the  $\mathbb{Z}_{2}$-permutation orbifolds 
$(G\times G )/\mathbb{Z}_{2}$, in which 
case the closure of the subset $\mathcal H$ under modular transformations
\begin{subequations} 
\begin{gather}
\chi _{(p - )} (\tau ) \equiv (\chi _{(p0)}  - \chi _{(p1)} )(\tau ) = \chi _p (2\tau ),\quad Z(2\tau ,2{\bar \tau }) \sim \sum_p {|\chi _p (2\tau )|^2 } \\ 
\mathcal{X}_{\widehat{(p \pm )}} (\tau ) \equiv \left( {\mathcal{X}_{\widehat{(p0)}}  \pm \mathcal{X}_{\widehat{(p1)}} } \right)(\tau )\\ 
\chi _{(p - )} (\tau  + 1) = \chi _{(p - )} (\tau ),\quad \mathcal{X}_{\widehat{(p \pm )}} (\tau  + 1) = T_p^{ - \tfrac{1}
{2}} \mathcal{X}_{\widehat{(p \mp )}} (\tau )\\ 
\chi _{(p - )} \left( { - \frac{1}
{\tau }} \right) = \sum_k {S_{pk} \mathcal{X}_{\widehat{(k + )}} (\tau )} \\ 
\mathcal{X}_{\widehat{(p + )}} \left( { - \frac{1}
{\tau }} \right) = \sum_k {S_{pk} \chi _{(k - )} (\tau )} \\ 
\mathcal{X}_{\widehat{(p - )}} \left( { - \frac{1}
{\tau }} \right) = \sum_k {P_{pk} } \mathcal{X} _{\widehat{(k - )}} (\tau ) 
\end{gather}
\end{subequations}
is easily read from  the modular transformations of the full set of 
orbifold characters in Ref. [13]. In this result, $\{ \chi_p(2\tau) 
\}$ are the diagonal untwisted characters, $\{ \mathcal{X}_{\widehat{(p 
\psi)}} (\tau)\}$ are the twisted characters, and the modular 
matrices $T,S,P$ are defined in the (untwisted) mother theory $G$ 
(see Eq. (5.8) and Ref. [13]).

For certain compactifications of the free-bosonic orbifolds at all 
prime $\lambda$, the work of Ref. [33] has previously identified a 
closely-related but not identical subset of characters which are 
closed under the modular 
subgroup $\Gamma_0(2)$ -- but not closed under the modular group itself. In our 
notation, this $\Gamma_0(2)$-covariant subset reduces for $H({\rm 
perm})=\mathbb{Z}_2$ to the particular subset  $\mathcal{S}=\{ \chi_{(p-)},\mathcal{X}_{_{\widehat{(p + )}} 
}\}$ of our modular-covariant subset $\mathcal H$ and, according to Eq. (5.9), the 
subset $\mathcal{S}$ is closed under  $\Gamma_0(2)$  for all the non-abelian 
orbifolds as well. Extensions of monster moonshine to $c=24k$ have 
also been considered in Ref. [34].



Between the ``elemental'' cases in Eqs. (5.6) and (5.7), I also mention the 
intermediate forms at $\beta=\lambda-1$ and $(\lambda-1)/\lambda$
\begin{subequations} 
\begin{gather}
\begin{split}
(\alpha '_c )^{13} {\hat \Lambda }^{(\lambda )}  =  - \frac{1}
{2}\int_F \frac{{d^2 \tau }}
{{(\operatorname{Im} \tau )^2 }}\Big\{& \lambda (2 - \lambda )Z(\tau ,{\bar \tau }) + \\ &+ (\lambda  - 1)\left[ Z(\lambda \tau ,\lambda {\bar \tau }) + \sum_{r = 0}^{\lambda  - 1} {Z\left( {\tfrac{{\tau  + r}}
{\lambda },\tfrac{{{\bar \tau } + r}}
{\lambda }} \right)}  \right] \Big\}
\end{split}\\ 
\begin{split}
(\alpha '_c )^{13} {\hat \Lambda }^{(\lambda )}  =  - \frac{1}
{{2\lambda }}\int_F \frac{{d^2 \tau }}
{{(\operatorname{Im} \tau )^2 }}\Big\{ & (\lambda ^2  - \lambda  + 1)Z(\tau ,{\bar \tau }) + \\ &+ (\lambda  - 1)\left[ Z(\lambda \tau ,\lambda {\bar \tau }) + \sum_{r = 0}^{\lambda  - 1} {Z\left( {\tfrac{{\tau  + r}}
{\lambda },\tfrac{{{\bar \tau } + r}}
{\lambda }} \right)} \right] \Big\}
\end{split} 
\end{gather}\end{subequations}
the scale of whose Hecke terms is close in spirit to the standard CFT 
prescription [8]. Note that the special case  $\lambda=2$ of Eq. (5.10a)

\begin{equation} 
	(\alpha '_c )^{13} {\hat \Lambda }^{(2)}  =  - \frac{1}
{{2\lambda }}\int_F {\frac{{d^2 \tau }}
{{(\operatorname{Im} \tau )^2 }}\left[ {Z(2\tau ,2{\bar \tau }) + Z\left( {\tfrac{\tau }
{2},\tfrac{{{\bar \tau }}}
{2}} \right) + Z\left( {\tfrac{{\tau  + 1}}
{2},\tfrac{{{\bar \tau } + 1}}
{2}} \right)} \right]} 
\end{equation}
coincides with the $\lambda=2$ case of Eq. (5.7). The physical 
spectral equivalence given in Eq. (3.11) can of course be applied to the twisted
contribution of any provisional form in (5.1), but the $\beta=\lambda-1$ case in 
Eq. (5.10a) is the only one for which the scale of the twisted 
contribution
\begin{equation}
	{\hat {\hat \Lambda }}^{(\lambda )}  =  - \frac{{\lambda  - 1}}
{2}\int_F {\frac{{d^2 \tau }}
{{(\operatorname{Im} \tau )^2 }}\sum_{r = 0}^{\lambda  - 1} {Z\left( {\tfrac{{\tau  + r}}
{\lambda },\tfrac{{{\bar \tau } + r}}
{\lambda }} \right)} }  =  - \frac{{\lambda  - 1}}
{2}\int_{{\hat F}_\lambda  } {\frac{{d^2 \tau }}
{{(\operatorname{Im} \tau )^2 }}\,Z(\tau ,{\bar \tau })} 
\end{equation}
transparently shows the (differently cut off) contribution of $\lambda-1$ ordinary 
closed strings. The new modular regions $\{\hat F_{\lambda}\}$ are 
defined in Eq. (3.11c). More generally, character analysis of 
intermediate cases such as those in  Eq. (5.10) would apparently require an 
auxiliary 
set of ordinary untwisted characters for the $Z(\tau, \bar\tau)$ term -- in addition to the new modular 
subset $\mathcal H$ of orbifold characters discussed above for the Hecke 
term.

\newpage

\section{One-Loop Diagrams with Insertions}

\subsection{Loops with a Cosmological Kernel}

Following the principles used to obtain the provisional forms of the orbifold cosmological 
constant, I have also worked out a corresponding set of provisional one-loop 
orbifold-string diagrams with an arbitrary number of untwisted 
tachyonic {\em insertions} at $ \alpha'_ck^2=-2$.  The twisted and untwisted contributions to loops of 
this type are depicted in Fig. 4, although the explicit labelling of the 
external lines in this figure corresponds to a later variant (see 
Eq. (6.13)) of the loops discussed here. For pedagogical purposes I 
will first present and discuss the final, physical form of these 
loops, postponing their formal particle-theoretic derivation until 
the latter part of this subsection.

The final result for this set  $\{{\hat L}_n^{(\lambda 
)}(\{k\},\beta)\}$ of loops with n insertions is structurally analogous to the orbifold 
cosmological constant $\hat \Lambda^{(\lambda)}(\beta)$:
\begin{subequations} 
\begin{gather}
\begin{split} {\text{(}}\alpha '_c )^{13} {\hat L}_n^{(\lambda )} (\{ k\} 
,\beta ) \equiv  g^n \int_F & \frac{{d^2 \tau }}
{{(2\operatorname{Im} \tau )^2 }}\Big{\{}  (2\lambda  - 1 - \beta (\lambda  + 1))\mathcal{L}_n (\tau ,{\bar \tau };\{ k\} ) + \\ 
 & +  \beta \left[ \mathcal{L}_n (\lambda \tau ,\lambda {\bar \tau };\{ k\} )  + \sum_{r = 0}^{\lambda  - 1} {\mathcal{L}_n \left( {\tfrac{{\tau  + r}}
{\lambda },\tfrac{{{\bar \tau } + r}}
{\lambda };\{ k\} } \right)}  \right] \Big{\}}
\end{split} \\ 
\begin{split}
\mathcal{L}_n (\tau ,{\bar \tau };\{ k\} ) \equiv  \,\,& (8\pi ^2 \operatorname{Im} \tau )^{ - 12} |\eta (e^{2\pi i\tau } )|^{ - 48} \times \\ 
& \quad \times (2i\operatorname{Im} \tau )\int_{G_\nu  (\tau )} \left( 
{\prod\limits_{l = 2}^n {d^2 \nu _l } } \right) \prod\limits_{i < j} {\chi (\nu _{ij} ,\tau ,{\bar \tau })^{2\alpha '_c k_i  \cdot k_j } } 
\end{split}\\ 
\chi (\nu ,\tau ,{\bar \tau }) = 2\pi e^{ - \pi \operatorname{Im} ^2 \nu /\operatorname{Im} \tau } \left| {\frac{{\theta _1 (\nu |\tau )}}
{{\theta '_1 (0|\tau )}}} \right|,\quad \nu _{ij}  = \nu _i  - \nu _j\\ 
G_\nu  (\tau ):\quad |\operatorname{Re} \nu _i | \leq \frac{1}
{2},\quad \operatorname{Im} \tau  \geq \operatorname{Im} \nu _i  \geq 
0,\quad i = 2, \ldots ,n\,\,. 
\end{gather}
\end{subequations}
In this construction, $F$ is the standard fundamental region and $\beta$ is the 
same parameter that appears in the orbifold cosmological constant $\hat 
\Lambda^{(\lambda)}(\beta)$ in Eq. (5.1). The role of the 
ordinary partition
 function $Z(\tau,\bar \tau)$ in the orbifold cosmological constant is now played by the modular-invariant 
densities $\{\mathcal L_{n}\}$ of the ordinary closed-string loops 
$L_{n}= L_{n}^{(1)}$, which are included in the result Eq. (6.1a) when $\lambda =1$:


\begin{equation} 
(\alpha '_c )^{13} {\hat L}_n^{(1)} (\{ k\} ) = g^n \int_F {\frac{{d^2 \tau }}
{{(2\operatorname{Im} \tau )^2 }}\,\mathcal{L}_n (\tau ,{\bar \tau };\{ k\} )}. 
\end{equation}
It follows that the $\beta$-dependent integrands in Eq. (6.1a) are modular-invariant for all prime $\lambda$.


I will call these provisional forms (6.1) the {\it loops with a cosmological kernel} because 
omission of the ``insertion factor'' in the second line of Eq. (6.1b) shows 
exactly the same provisional forms (5.1) of the orbifold cosmological constant
\begin{equation} 
{\hat L}_n^{(\lambda )} (\{ k\} ,\beta ) \to {\hat\Lambda}^{(\lambda )} (\beta )
\end{equation}
as the ``kernel'' of each loop. This property is well-known (see e.f. 
Ref. [23]) for the 
ordinary closed-string loops at $\lambda=1$, but we will see in the 
following subsection that the insertions and hence the kernel can be changed
for the orbifolds with $\lambda \geq 2$.

Following the discussion in Sec. 5, I mention two special cases of 
the loops with a cosmological kernel:
\begin{subequations} 
\begin{gather}
\beta  = 0:\quad (\alpha '_c )^{13} {\hat L}_n^{(\lambda )} (\{ k\} ) = 
g^n (2\lambda  - 1)\int_F {\frac{{d^2 \tau }}
{{(2\operatorname{Im} \tau )^2 }}\,\mathcal{L}_n (\tau ,{\bar \tau };\{ k\} )} \\ 
{\hat L}_n^{(\lambda )} (\{ k\} ) = (2\lambda  - 1)L_n (\{ k\} ) 
\end{gather}
\end{subequations}

\begin{subequations} 
\begin{gather}
\begin{split}
\beta  = \lambda  - 1:\quad & (\alpha '_c )^{13} {\hat L}_n^{(\lambda )} (\{ k\} ) =  g^n \int_F \frac{{d^2 \tau }}
{{(2\operatorname{Im} \tau )^2 }}\Big{\{} \lambda (2 - \lambda )\mathcal{L}_n (\tau ,{\bar \tau };\{ k\} )+ \\ &+ (\lambda  - 1)\left[ {\mathcal{L}_n (\lambda \tau ,\lambda {\bar \tau };\{ k\} ) + \sum_{r = 0}^{\lambda  - 1} {\mathcal{L}_n \left( {\tfrac{{\tau  + r}}
{\lambda },\tfrac{{{\bar \tau } + r}}
{\lambda }};\{ k\} \right)} } \right] \Big{\}}  
\end{split} \\ 
\int_F {\frac{{d^2 \tau }}
{{(2\operatorname{Im} \tau )^2}}\sum_{r = 0}^{\lambda  - 1} {\mathcal{L}_n \left( {\tfrac{{\tau  + r}}
{\lambda },\tfrac{{{\bar \tau } + r}}
{\lambda };\{ k\} } \right) = \int_{{\hat F}_\lambda  } {\frac{{d^2 \tau }}
{{(2\operatorname{Im} \tau )^2 }}} } } \,\mathcal{L}_n (\tau ,{\bar \tau };\{ k\} ). 
\end{gather}
\end{subequations}
The provisional form in Eq. (6.4) is again the case favored 
by Occam's razor -- showing contributions to the orbifold loop of 
$\lambda$ ordinary strings from the untwisted sector and $\lambda-1$ 
ordinary strings from the twisted sectors. Beyond this case, the form in 
Eq. (6.5) is again 
the only case for which the scale of the twisted contribution to the 
physical loops transparently shows the (differently cut off) contribution of 
$\lambda-1$ ordinary strings -- the physical spectral-equivalence 
identity in Eq. (6.5b) 
following by cut and paste as discussed in Subsec. 3.3.


I turn now to sketch the formal derivation of the orbifold loops 
(6.1) from their 
particle-theoretic contributions, which will also help to understand 
the insertions in these loops. For this discussion we will need the well-known 
{\it trace forms}  of the ordinary closed-string loops, as they are 
obtained by sewing [34,12] of the ordinary closed-string trees:
\begin{subequations} 
\begin{gather}
\begin{align}
(\alpha '_c )^{13} L_n (\{ k\} )_N & = g^n \int {(d^{26} J(0))\{ Tr(\gamma (k_n ,1,1)D \ldots D\gamma (k_1 ,1,1) +  \cdots } \} \\ 
 & = g^n \int_{F_N } {\frac{{d^2 \tau }}
{{(2\operatorname{Im} \tau )^2 }}\,\mathcal{L}_n (\tau ,{\bar \tau };\{ k\} )} 
\end{align}\\ 
\begin{split} 
\mathcal{L}_n (\tau ,{\bar \tau };\{ k\} ) = & (2\operatorname{Im} \tau )^2  \int {(d^{26} J(0))(2\pi )^n \int_{\tilde G_{^\nu  } (\tau )} {\prod\limits_{i = 2}^n {d^2 \nu _i } } } \times \\
& \times \Big{\{}  Tr \big{(}\gamma (k_n ,1,1)\xi (\nu _n ,{\bar \nu }_n ;K_{n - 1}  - k)\gamma (k_{n - 1} ,1,1)\\
& \quad\quad \cdot \xi (\nu _{n - 1,n} ,{\bar \nu }_{n - 1,n} ;K_{n - 2}  - k) \ldots \gamma (k_1 ,1,1)\xi (\nu _2 ,{\bar \nu }_2 ;k)\\ 
& \quad\quad \cdot e^{ - 2\pi \operatorname{Im} \tau (L(0) + {\bar L}(0) - 2)} e^{2\pi i\operatorname{Re} \tau (L(0) - {\bar L}(0))} \big{)} +  \cdots \Big{\}} 
\end{split}\\ 
\xi (\nu ,{\bar \nu };k) = e^{2\pi i(\nu L(0) - {\bar \nu }{\bar L}(0))} 
,\,\,\quad K_i  = \sum_{j = 1}^i {k_j } ,\,\,\quad J(0) = \sqrt {\alpha '_c } k\\ 
\tilde G_\nu  (\tau ):\quad \operatorname{Im} \tau  \geq \operatorname{Im} \nu _2  \geq  \ldots  \geq \operatorname{Im} \nu _n  \geq 0,
\quad |\operatorname{Re} \nu _i | \leq \frac{1}
{2} \,.
\end{gather}
\end{subequations}
This is the naive  or {\it particle-theoretic} form of the ordinary loops, 
where $F_{N}$ is the naive modular region in Eq. (3.3c),  $D$ is the untwisted 
propagator in Eq. (3.8c), $\gamma(k)$ is the ordinary closed-string vertex operator
 at $ \alpha'_ck^2=-2$ and 
the ellipses denote symmetrization with respect to the external 
momenta $\{k\}$. 


Including the contributions of both the untwisted and twisted sectors, 
the corresponding trace form of the naive or {\it particle-theoretic} 
orbifold loop is then easily written down
\begin{subequations} 
\begin{gather}
\begin{split}	(\alpha '_c )^{13} {\hat L}_n^{(\lambda )} (\{ k\} )_N & = g^n \Big{\{}  \int {(d^{26} J(0))} Tr(\tilde \gamma (k_n ,1,1)\tilde D \ldots \tilde D \tilde \gamma (k_1 ,1,1)) + \\ & + (\lambda  - 1)\int {(d^{26} {\hat J}(0))} Tr(\hat{\hat g} (k_n ,1,1){\hat D} \ldots {\hat D}\hat{\hat g}(k_1 ,1,1) +  \cdots  \Big{\}}
\end{split}\\ 
\tilde \gamma _{IJ} (k,{\bar z},z) \equiv \gamma _I (k, {\bar z},z)\delta 
_{IJ} ,\quad\,\, I,J = 0,1, \ldots ,\lambda  - 1\\ 
\tilde D_{IJ} \equiv D[L^I_0(0), {\bar L}^I_0(0)] \delta_{IJ}\\ 
\hat{\hat g}(k,{\bar z},z) \equiv \,{\hat g}\left( {\mathcal{T} = tT \to 
\sqrt {\alpha '_c } k,{\bar z},z} \right),\quad\,\, \alpha '_c k^{2}=\,-2. 
\end{gather}
\end{subequations}
where $D$ and $\hat D$ are the untwisted and twisted propagators in Eq. 
(3.8c). In this expression,
 I have chosen the untwisted vertex operator $\tilde 
\gamma$ to describe untwisted tachyonic emission from the untwisted 
propagator $\tilde D$. Both structures $\tilde \gamma$ and $\tilde D$ are $\lambda\times\lambda$ 
matrices, with a diagonal entry for each copy $I$ in the untwisted 
sector. The trace in the first term of Eq. (6.7a) is defined to include 
the trace over $\{I,J\}$ space of the matrix product of these operators.
For untwisted tachyonic emission from the twisted propagator $\hat D$, 
I have for simplicity chosen the reduced form $\hat{\hat g}$ of the twisted vertex 
operator, which masks the corresponding  $\lambda\times\lambda$ 
matrix structure of the twisted vertex operator $\hat g$ (see Eqs. (2.11) and (2.12)).


The trace-evaluated form of the sewn orbifold loop in Eq. (6.7) is as 
follows
\begin{multline} 
	(\alpha '_c )^{13} {\hat L}_n^{(\lambda )} (\{ k\} )_N  = \int_{F_N } {\frac{{d^2 \tau }}
{{(2\operatorname{Im} \tau )^2 }}} \Big{\{} \lambda \mathcal{L}_n (\tau ,{\bar \tau };\{ k\} ) + \\ + (\lambda  - 1)\sum_{r = 0}^{\lambda  - 1} {\mathcal{L}_n \left( {\tfrac{{\tau  + r}}
{\lambda },\tfrac{{{\bar \tau } + r}}
{\lambda };\{ k\} } \right)}  \Big{\}}
\end{multline}
where $\{\mathcal L_{n}\}$ are the untwisted loop densities, and one also finds the formal
integration identities
\begin{subequations} 
\begin{gather}
\int_{F_N } {\frac{{d^2 \tau }}
{{(\operatorname{Im} \tau )^2 }}\left\{ {\sum_{r = 0}^{\lambda  - 1} {\mathcal{L}_n \left( {\tfrac{{\tau  + r}}
{\lambda },\tfrac{{{\bar \tau } + r}}
{\lambda };\{ k\} } \right) - \mathcal{L}_n (\tau ,{\bar \tau };\{ k\} )} } \right\} = 0} \\ 
\int_{F_N } {\frac{{d^2 \tau }}
{{(\operatorname{Im} \tau )^2 }}\{ \mathcal{L}_n (\lambda \tau ,\lambda {\bar \tau };\{ k\} ) - \lambda \mathcal{L}_n (\tau ,{\bar \tau };\{ k\} )\} }  = 0 
\end{gather}
\end{subequations}
which correspond respectively to the spectral equivalences in Eqs. (3.11a) 
and (4.8) for the orbifold
cosmological constant. In particular, Eq. (6.9a) is a cut-and-paste identity (see 
Subsec. 3.3), and the identity in Eq. (6.9b)  
follows from the trace form (6.6c) of the densities $\{\mathcal L_{n}\}$
 by steps analogous to those shown in Eq. (4.9).

I should emphasize that the untwisted vertex operator $\tilde\gamma$ 
in Eq. (6.7) 
couples identically to all the untwisted internal strings, which gives the 
factor $\lambda$ in the first term of Eq. (6.8). This property is also 
noted in the
 Appendix, where the tree amplitudes of the 
vertex $\tilde\gamma$ are evaluated in terms of the ordinary 
closed-string trees. I have explicitly checked 
the twisted contribution of Eq. (6.7) to Eq. (6.8) only for the case $\lambda=2$, but 
the result is certainly correct for all prime $\lambda$ because Eq. 
(6.9a) shows that the trace-evaluated contribution of each twisted sector
is equivalent to that of an ordinary closed string.


Thanks to the specific choice of emission vertices in Eq. (6.7), the reader will 
note that (up to a scale) all these trace-evaluated particle-theoretic results can be obtained
 by the simple map
\begin{equation} 
	Z(\tau ,{\bar \tau }) \,\to \,\mathcal{L}_n (\tau ,{\bar \tau };\{ k\} )
\end{equation}
from the corresponding results above for the orbifold cosmological constant. 
It is then straightforward to write down the loop analogues of Eqs. 
(5.3) 
and (5.4) and divide by $SL(2,\mathbb{Z})$ to obtain the final result 
in Eq. (6.1) for the 
physical loops with a cosmological kernel.


\subsection{Emissions of a Representative Tachyon}

The orbifold-string loops (6.1) with a cosmological kernel were easy to understand in 
analogy with the orbifold cosmological constant but, as mentioned above, 
one can also construct modified orbifold loops which do not have a 
cosmological kernel when $\lambda \geq 2$.

I find it instructive to consider this topic first in a broader setting, 
beginning with the following more general set of manifestly modular-invariant integrals:
\begin{subequations} 
\begin{multline}
  (\alpha '_c )^{13} {\hat L}_n^{(\lambda )} (\{ k\} ;\beta ,\rho ,\eta ) = g^n \int_{F} {\frac{{d^2 \tau }}
{{(2\operatorname{Im} \tau )^2 }}} \Big\{ A\mathcal{L}_n (\tau ,{\bar \tau };\{ k\} ) + \quad\quad \\
  \quad\quad + B \Big[\mathcal{L}_n (\lambda \tau ,\lambda {\bar \tau };\{ k\} ) + \sum_{k = 0}^{\lambda  - 1} {\mathcal{L}_n (\tfrac{{\tau  + r}}
{\lambda },} \tfrac{{{\bar \tau } + r}}
{\lambda };\{ k\} )\Big] \Big\}  
\end{multline} 
\begin{equation} 
A = \rho \lambda  + \eta (\lambda  - 1) - \rho \beta (\lambda  + 1),\quad B = \rho \beta. 
\end{equation}
\end{subequations}
In this expression, $\{\mathcal L_{n}\}$ are the same ordinary, 
untwisted string densities 
and the parameter $\beta $ is the same as above, but 
 $\rho$ and $\eta$ are new complex parameters.
Following the procedure outlined above for the loops with a 
cosmological kernel, these integrals can in fact be obtained 
by the map (6.10) and  the corresponding loop-analogues of 
the spectral-equivalence identities (5.3) from the following naive form of the 
integrals
\begin{multline} 
(\alpha '_c )^{13} {\hat L}_n^{(\lambda )} (\{ k\} ,\rho ,\eta )_N  = \int_{F_N } \frac{{d^2 \tau }}
{{(2\operatorname{Im} \tau )^2 }} \Big\{ \rho \lambda \mathcal{L}_n (\tau ,{\bar \tau };\{ k\} ) + \\ 
+ \eta (\lambda  - 1) \sum_{r = 0}^{\lambda  - 1} {\mathcal{L}_n (\tfrac{{\tau  + r}}
{\lambda },} \tfrac{{{\bar \tau } + r}}
{\lambda };\{k\})\Big\}.
\end{multline}
In particular,  the parameter choice  $\gamma  = (\rho \beta /\eta (\lambda  - 
1))$   is necessary to obtain the modular-invariant 
integrands of Eq. (6.11) from the non-invariant integrands of Eq. (6.12).
It is not clear that this naive form is always 
associated to an operator sewing, but we can be confident in certain 
cases: In the first place, we find for arbitrary values of the 
parameters and $\lambda=1$ that both Eqs. (6.11) and (6.12) reduce to $\rho$ times 
the ordinary parameter-independent untwisted string loops. Second, we recognize 
the case $\rho=\eta=1$ as the physical and naive forms respectively  
of the orbifold-loops with a cosmological kernel (see Eqs. (6.1) 
and (6.8)).

Another case of particular interest is the parameter choice  $\rho 
=1/\lambda, \eta=1$, which I will call the {\it orbifold loops of type 
$I$}. These are the 
$I$-independent  loops  for n emissions of a 
representative untwisted tachyon of type $I$:
\begin{subequations} 
\begin{multline}
(\alpha '_c )^{13} {\hat L}_n^{(\lambda )} (\{ k\} )_{NI}  = g^n 
\int_{F_{N}} {\frac{{d^2 \tau }}
{{(2\operatorname{Im} \tau )^2 }}} \Big\{ \mathcal{L}_n (\tau ,{\bar \tau };\{ k\} ) + \\
+(\lambda  - 1)\sum_{r = 0}^{\lambda  - 1} {\mathcal{L}_n (\tfrac{{\tau  + r}}
{\lambda },} \tfrac{{{\bar \tau } + r}}
{\lambda };\{k\}) \Big\} 
\end{multline} 
\begin{multline}
   \to (\alpha '_c )^{13} {\hat L}_n^{(\lambda )} (\{ k\} ,\beta )_I  = g^n \int_{F} {\frac{{d^2 \tau }}
{{(2\operatorname{Im} \tau )^2 }}} \Big\{ (\lambda  + \frac{\beta }
{\lambda }(\lambda  + 1))\mathcal{L}_n (\tau ,{\bar \tau };\{ k\} ) +  \\ 
   + \beta \Big[\mathcal{L}_n (\lambda \tau ,\lambda {\bar \tau };\{ k\} ) + \sum_{r = 0}^{\lambda  - 1} {\mathcal{L}_n (\tfrac{{\tau  + r}}
{\lambda },} \tfrac{{{\bar \tau } + r}}
{\lambda };\{k\})\Big] \Big\} 
\end{multline} 
\begin{equation} 
I = 0,1, \ldots ,\lambda  - 1
\end{equation}
\end{subequations}
I have been careful to uniformize this notation -- and the 
depiction of these loops in Fig. 4 -- with the corresponding tree 
graphs $\{\hat A_{nI}^{(\lambda)}\}$
in Eq. (2.12) and Fig. 1.  Comparing with the trace-evaluated result 
(6.8) for 
the naive loops with a cosmological kernel, we see that the internal twisted 
contribution to Eq. (6.13a) is here unchanged -- but the kernel of the new 
loops is {\it not 
cosmological} for $\lambda \geq 2$ because the first term now 
shows only the contribution of a {\it single} untwisted internal closed string.
 The  trace form of the naive loops $\{{\hat L}_n^{(\lambda)}(\{k\})_{NI}\}$ is then
simply obtained by substitution of the type-$I$ emission vertices 
(see Eqs. (2.11), (2.12) and (6.7b))
\begin{equation}
 \tilde \gamma \rightarrow 
\gamma_{I},\quad\quad \hat{\hat g} \rightarrow {\hat g}_{I}
\end{equation}  
in the trace form (6.7a) of the loop (and omission of the previously-implied trace 
over $\{I,J\}$ space in the untwisted term). Use of the type-$I$ emission vertices  does not change the 
internal contribution of the twisted sectors, but restricts the internal untwisted 
contribution to that of {\it type $I$ alone} -- thereby completing our 
identification of these loops. Because these loops and the 
corresponding trees of type $I$ are independent of $I$, they may be 
used (in lieu of moding by permutations of $\{I\}$) at any fixed $I$ to 
describe multiple emissions of a representative tachyon.


\begin{figure}[htbp]
	\centering
		\includegraphics[angle=0,width=0.5\columnwidth]{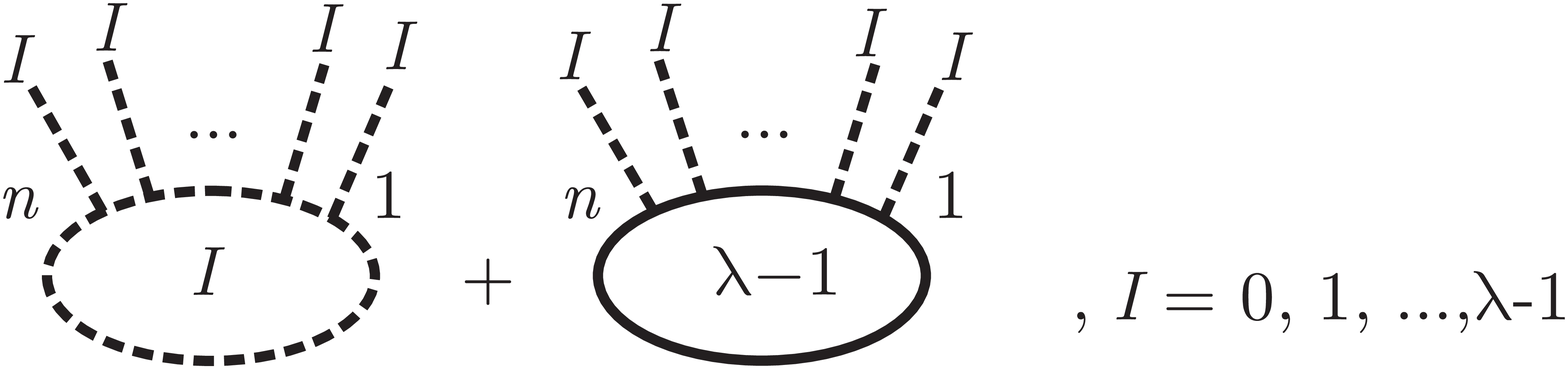}
	\caption{Emissions of a representative tachyon.}
	\label{fig:emissions}
\end{figure}

In parallel to  Eqs. (5.6) and (6.4), I finally\ mention the simplest case $\beta=0$ of the physical loops 
(6.13b)
\begin{equation} 
	(\alpha '_c )^{13} {\hat L}_n^{(\lambda )} (\{ k\} )_I  = g^n \lambda \int_F {\frac{{d^2 \tau }}
	{{(2\operatorname{Im} \tau )^2 }}} \,\mathcal{L}_n (\tau ,{\bar \tau };\{ k\} )
\end{equation}
which clearly shows  a single ordinary closed-string 
 contribution from the untwisted sector and $\lambda-1$ ordinary 
 contributions from the twisted sectors.
 
\newpage
\section{Conclusions}

Historically, conformal field theory and orbifold theory arose from 
string theory, and these five papers  have outlined my ideas on 
closing the 
 circle by constructing new physical string systems at higher central 
 charge from the 
conformal field theory of orbifolds of permutation-type.

Many extensions of these ideas are indicated for future work. In 
particular,  Ref. [31] and the present paper have so far considered only the 
very simplest of the new orbifold-string systems  at the 
interacting level. These particular constructions allowed us to 
study the interacting systems in a context where, as we have seen, the 
physical spectra of the twisted strings are closely related to 
untwisted string spectra. Hopefully, these examples will then 
serve as prototypes for the study of the more general
interacting orbifold-string systems of permutation-type, whose spectra are 
in fact generically-inequivalent [30] to untwisted systems.


Consider for example the {\it generalized} bosonic 
$\mathbb{Z}_{2}$-permutation 
 orbifolds $U(1)^{52}/ H_{+}$ [28-30] with sectors labelled by elements 
 of $ H_{+}= \mathbb{Z}_{2}\times H $:
\begin{subequations}
\begin{gather}
(1;\,\,\, \tau_{+}\times\omega_{2}),\quad \omega_{2}^{2}=1
\end{gather}
\begin{gather}
(1,\,\,\omega_{3},\,\,\omega_{3}^{2};\,\,\,\tau_{+},\,\,\tau_{+}\times\omega_{3},\,\,\tau_{+}\times\omega_{3}^{2}),
\quad \omega_{3}^{3}=1
\end{gather}
\begin{gather}
(1,\,\,\omega_{4}^{2};\,\,\,\tau_{+}\times\omega_{4},\,\,\tau_{+}\times\omega_{4}^{3}), \quad\omega_{4}^{4}=1.
\end{gather}
\end{subequations}
Here $\tau_{+}$ permutes the two copies of untwisted $U(1)^{26}$,  and 
$\{\omega_{n}\}\in H$ are extra automorphisms of 
order n which act 
uniformly on each copy. The low-order examples given here are easily extended to 
all n.  All the sectors of this set of orbifolds 
are closed strings at $\hat c=52$, the entries before and after the 
semicolon corresponding to untwisted and twisted sectors respectively. 
The extended Virasoro generators of all twisted open and closed
 $\hat c=52$ strings were 
constructed in Ref. [30], and I remind that the $\hat c=52$ string 
spectra are {\it not equivalent} to untwisted spectra when $H$ has elements 
of order greater than two. This statement includes the twisted 
$\hat c=52$ open-string spectra of the general bosonic orientation 
orbifolds 
$U(1)^{26}/(\mathbb{Z}_{2}(w.s.)\times H)$ [27-31].

One may further consider the generalized cyclic 
permutation orbifolds $U(1)^{26\lambda}/H_{+}$ [28,30] with $H_{+}=\mathbb{Z}_{\lambda}\times H$ at 
$\hat c=26\lambda$ for 
higher prime $\lambda$, all 
of whose twisted tree graphs should have the form in Eq. (2.9) with the 
universal value (2.5b) of the quantity $\hat a_{\lambda}$ in the twisted 
propagators. The twisted vertex operators for these orbifolds can be read as special cases of the general twisted 
bosonic vertex operators in Ref. [19]. For all these cases,
an equivalent but {\it unconventionally-twisted} $c=26$ description of the 
spectrum is easily written down using a simple variant (see Refs. 
[13,30] and 
Subsec. 2.3) of the {\it orbifold-induction procedure} with
\begin{subequations}
\begin{gather}
J(M+\tfrac{\lambda n(r)}{\rho(\sigma)}) \equiv \,\,
\hat J(m+\tfrac{n(r)}{\rho(\sigma)}+\tfrac{r}{\lambda}),\quad  M\equiv 
\lambda m +r
\end{gather}
\begin{gather}
r=0,1\ldots,\lambda -1, \quad n(r)\in (0,1,\ldots, \rho(\sigma)-1).
\end{gather}
\end{subequations}
This map and the unhatted $c=26 $ modeing $(M+\tfrac{\lambda 
n(r)}{\rho(\sigma)}) $ tells us that the physical spectra of these systems are again {\it 
inequivalent} to 
untwisted spectra cases unless $(\lambda n(r)/\rho(\sigma))\in \mathbb{Z}$ 
for all ``H-fractions'' $\{n(r)/\rho(\sigma)\}$. (The quantities 
$\{n(r)\}$ and $\rho(\sigma)$ are the spectral indices [15,17,19,21,28] and order 
respectively of any element $\omega(\sigma)$ of H.)

Although I will not give the details here, one can in fact write down a 
generalized orbifold-induction procedure for every sector of the 
{\it generalized} permutation orbifold $U(1)^{26K}/(H(\text{perm})\times 
H)$ [18,19,28] at $\hat c=26K$, using a distinct map for each cycle in each element of 
the general permutation group $H(\text{perm})$ on K elements. On this basis, one sees that the physical 
spectra of
these general systems are related to untwisted string spectra {\it only}
for the special cases when $(f_{j}(\sigma) n(r)/\rho(\sigma))\in \mathbb{Z}$ for all cycle 
lengths $\{f_{j}(\sigma)\}$ and H-fractions $\{n(r)/\rho(\sigma)\}$ of 
each sector. 
The simple subset of ``ordinary'' permutation orbifolds 
$U(1)^{26K}/H(\text{perm})$ at $\hat c=26K$ satisfy 
these conditions with $(n(r)/\rho(\sigma))=1$, so   
(consistent with our work in this paper and in Refs. [28-30]) the physical spectra 
of all the corresponding ``ordinary'' permutation-orbifold strings are closely related 
to untwisted string spectra. Beyond these special cases, one expects [28-31] the 
vast majority of generalized permutation orbifolds to describe a great 
variety of new orbifold-string theories, yet to be studied at the interacting level.

Other directions, including partial orbifoldizations, partial compactifications, twisted B fields 
and the corresponding extensions to the superstring 
orbifolds of permutation-type are 
sketched in Ref. [28] and the introduction to this paper.

\section*{Acknowledgements}

For helpful discussions and encouragement, I thank L. 
Alvarez-Gaum$\acute{e}$, K. Bardakci, I. Brunner, J. de Boer, D. Fairlie, O. 
Ganor, E. Gimon, C. Helfgott, E. Kiritsis, R. Littlejohn, S. 
Mandelstam, J. McGreevy, N. Obers, A. Petkou, E. Rabinovici, V. Schomerus, K. 
Schoutens, C. Schweigert and E. Witten. 

\appendix
\section{The Trees of the Vertex $ \mbox{\boldmath{$\tilde{\gamma}$}}$ }

The untwisted emission vertex $\tilde\gamma$ 
and propagator $\tilde D$ are defined
in Eqs. (6.7b) and (6.7c) of the text. Both structures are 
$\lambda\times\lambda$ matrices with a diagonal entry for each 
closed-string copy 
$I=0,\ldots\lambda-1$ in the untwisted sector of the orbifold. In accord with their  
loop contribution in Eq. (6.7a), the 
tree amplitudes $\{\tilde A_n^{(\lambda)}\}$ of the tilde system include 
 a trace over matrix multiplication of these operators -- 
which guarantees that the vertex $\tilde\gamma$ couples identically to all $\lambda$ 
copies in the untwisted sector. 
This is clearly seen in the explicit evaluation of these trees
 \begin{subequations} 
\begin{gather}
	\widetilde{|0\rangle } \equiv |0\rangle _0  \otimes  \cdots  \otimes |0\rangle _{\lambda  - 1} \\ 
	\begin{align}
	\widetilde{|k\rangle }_{IJ} & 
	\equiv \mathop {\lim }\limits_{z \to 0} \tilde \gamma _{IJ} (k,{\bar z},z)\widetilde{|0\rangle}\\ 
	 & = \delta _{IJ} |0\rangle _0  \otimes  \cdots  \otimes |k\rangle _I  \otimes  \cdots  \otimes |0\rangle _{\lambda  - 1}
	 \end{align}\\ 
	 \begin{align}
	 \tilde A_n^{(\lambda )} (\{ k\} ) & \equiv Tr(\widetilde{\langle k_n |}
	 \tilde \gamma (k_{n - 1} ,1,1)\tilde D \ldots \tilde D\tilde \gamma (k_2 ,1,1)\widetilde{|k_1 \rangle } 
	 +  \ldots ) \\
	 & = \sum_I {{}_I\langle k_n |\gamma _I (k_{n - 1} ,1,1)D_I  \ldots D_I \gamma _I (k_2 ,1,1)|k_1 \rangle _I  
	 +  \ldots } \\ 
	   & = \lambda A_n (\{ k\} )\,=\, Tr(\thickone_\lambda   \ldots \thickone_\lambda  )A_n (\{ k\} )
	   \end{align}\\ 
	   \begin{align}
	   A_n (\{ k\} ) & = \tilde A_n^{(1)} (\{ k\} )  \\ 
	   & = \langle k_n |\gamma (k_{n - 1} ,1,1)D \ldots D\gamma (k_2 ,1,1)|k_1 \rangle  +  \ldots  
	   \end{align} 
\end{gather}
\end{subequations}
where $\{A_{n}\}$ are the trees of the ordinary closed-string.


\begin{thebibliography}{}

\bibitem{BH} 
K.~Bardakci and M.~B. Halpern, ``New dual quark models,'' {\em Phys. Rev.} {\bf 
  D3} (1971) 2493.
  
\bibitem{CFT}
M.~B. Halpern, ``The two faces of a dual pion-quark model,'' {\em Phys. Rev.} 
  {\bf D4} (1971) 2398; R.~Dashen and Y.~Frishman, ``Four-fermion interactions and scale invariance,'' 
{\em Phys. Rev.} {\bf D11} (1975) 2781;  M.~B. Halpern, ``Quantum `solitons' which are SU(N) fermions,'' {\em Phys. 
Rev.} {\bf D12} (1975) 1684; V.~G. Knizhnik and A.~B. Zamolodchikov, ``Current algebra and Wess-Zumino model in
   two dimensions,'' {\em Nucl. Phys.} {\bf B247} (1984) 83;  M.~B. Halpern and E.~Kiritsis, ``General Virasoro Construction on 
affine g,''  {\em Mod. Phys. Lett.}  {\bf A4} (1989) 1373; M.~B.Halpern, E.~Kiritsis, N.~A. Obers and K.~Clubok, ``Irrational 
conformal field theory,'' {\em Phys. Rep.} {\bf 265} (1996) 1, hep-th/9501144. 


\bibitem{2faces2}
M.~B. Halpern and C.~B. Thorn, ``The two faces of a dual pion-quark model 
II. Fermions and other things,'' {\em Phys. Rev.} {\bf D4} (1971) 3084.

\bibitem{Corr}
E.~Corrigan and D.~B. Fairlie, ``Off-shell states in dual resonance 
theory,''  {\em Nucl. Phys.}{\bf B91} (1975) 527; W.~Siegel, ``Strings with dimension-dependent intercept,''
 {\em Nucl. Phys.} {\bf B109} (1976) 244.

\bibitem{Lep}
J.~Lepowsky and R.~L. Wilson, ``Construction of the affine Lie 
algebra $A_{1}^{(1)}$,'' {\em Comm. Math. Phys.} {\bf 62} (1978) 43.

\bibitem{Orb}
L.~Dixon, J.~Harvey, C.~Vafa and E.~Witten, ``Strings on orbifolds,'' 
{\em Nucl. Phys.} {\bf B261} (1985) 678; ``Strings on orbifolds II,'' 
{\em Nucl. Phys.} {\bf B274} (1986) 285. L.~Dixon, D.~Friedan, E.~Martinec and S.~Shenker, `` The conformal 
field theory of orbifolds,'' {\em Nucl. Phys.} {\bf B282} (1987) 13. S.~Hamidi and C.~Vafa, ``Interactions on orbifolds,'' {\em Nucl. 
Phys.} {\bf B279} (1987) 465. J.~K. Freericks and M.~B. Halpern, ``Conformal deformation by the currents 
of affine g,'' {\em Ann. Phys.} {\bf 188} (1988) 258.

\bibitem{DVVV}
 R.~Dijkgraff, C.~Vafa, E.~Verlinde and H.~Verlinde, `` The operator 
algebra of orbifold models,'' {\em Comm. Math. Phys.} {\bf 123} (1989) 485.

\bibitem{Klemm}
A.~Klemm and M.~G. Schmidt, ``Orbifolds by cyclic permutations of 
tensor-product conformal field theories, '' {\em Phys. Lett.} {\bf 
B245} (1990) 53.

\bibitem{String}
G.~Veneziano, ``Construction of a crossing-symmetric Regge-behaved 
amplitude for linearly rising trajectories,'' {\em Nuovo Cimento} {\bf 
57A} (1968) 190. 

\bibitem{closed}
 M.~A. Virasoro, ``Alternative constructions of crossing-symmetric 
amplitudes with Regge behavior,'' {\em Phys. Rev.} {\bf 177} (1969) 
2309; J.~Shapiro, ``Electrostatic analogue for the Virasoro model,'' {\em Phys. 
Lett.} {\bf B33} (1970) 361.

\bibitem{stan}
S.~Mandelstam ``Dual resonance models,'' {\em Phys. Rep.} {\bf 13} (1974) 259.

\bibitem{GSW}
M.~B. Green, J.~H. Schwarz and E.~Witten, ``Superstring theory,''  
Cambridge University Press, 1987.


 \bibitem{Christ}
L.~Borisov, M.~B. Halpern, and C.~Schweigert, ``Systematic approach to cyclic orbifolds,''
 {\em Int. J. Mod. Phys.} {\bf A13} (1998) 125, hep-th/9701061. 
 
\bibitem{OVME} 
J.~Evslin, M.~B. Halpern, and J.~E. Wang, ``General {Virasoro} construction on 
  orbifold affine algebra,'' {\em Int. J. Mod. Phys.} {\bf A14} (1999) 
  4985, hep-th/9904105. 
 
\bibitem{Dual} 
J.~de~Boer, J.~Evslin, M.~B. Halpern, and J.~E. Wang, ``New duality 
  transformations in orbifold theory,'' {\em Int. J. Mod. Phys.} {\bf A15} 
  (2000) 1297, hep-th/9908187. 

 \bibitem{Coset} 
J.~Evslin, M.~B. Halpern, and J.~E. Wang, ``Cyclic coset orbifolds,'' {\em Int. 
  J. Mod. Phys.} {\bf A15} (2000) 3829,  hep-th/9912084. 
 
\bibitem{More} 
M.~B. Halpern and J.~E. Wang, ``More about all current-algebraic orbifolds,'' 
  {\em Int. J. Mod. Phys.} {\bf A16} (2001) 97, hep-th/0005187. 
 
\bibitem{Big} 
J.~de~Boer, M.~B. Halpern, and N.~A. Obers, ``The operator algebra and twisted 
  {KZ} equations of {WZW} orbifolds,'' {\em J. High Energy Phys.} {\bf 10} (2001) 011, 
 hep-th/0105305. 
 
\bibitem{Big'} 
M.~B. Halpern and N.~A. Obers, ``Two large examples in orbifold theory: 
Abelian orbifolds and 
  the charge conjugation orbifold on $su(n)$,'' {\em Int. J. Mod. Phys.} {\bf A17} (2002) 3897, hep-th/0203056.

\bibitem{Fab} 
M.~B. Halpern and F.~Wagner, ``The general coset orbifold action,'' {\em Int. J. Mod. Phys.}
  {\bf A18} (2003) 19, hep-th/0205143. 

\bibitem{Perm}
M.~B. Halpern and C.~Helfgott, ``Extended operator algebra and reducibility in the WZW permutation orbifolds,''
    {\em Int. J. Mod. Phys.} {\bf A18} (2003) 1773,  hep-th/0208087. 
    
\bibitem{So2n}
O.~Ganor, M.~B. Halpern, C.~Helfgott and N.~A. Obers, ``The outer-automorphic 
WZW orbifolds on $so (2n)$, including
    five triality orbifolds on $so(8)$,'' {\em J. High Energy Phys.} {\bf 0212} (2002) 019, hep-th/0211003. 

\bibitem{Geom}
J.~de~Boer, M.~B. Halpern and C.~Helfgott, ``Twisted Einstein tensors and 
orbifold geometry,"
 {\em Int. J. Mod. Phys.} {\bf A18} (2003) 3489, hep-th/0212275.


\bibitem{Orient1}
M.~B. Halpern and C.~Helfgott, ``Twisted open strings from closed strings: 
The WZW orientation orbifolds,"
 {\em Int. J. Mod. Phys.} {\bf A19} (2004) 2233, hep-th/0306014.

\bibitem{Orient2}
M.~B. Halpern and C.~Helfgott, ``On the target-space geometry of the 
open-string orientation-orbifold sectors," {\em Ann. of Phys.}
  {\bf 310} (2004) 302, hep-th/0309101. 

\bibitem{Basic}
M.~B. Halpern and C.~Helfgott, ``A basic class of twisted open WZW strings,"
{\em Int. J. Mod. Phys.} {\bf A19} (2004) 3481, hep-th/0402108.

\bibitem{Gentwopen}
M.~B. Halpern and C.~Helfgott, ``The general twisted open WZW string,'' 
{\em Int. J. Mod. Phys.} {\bf A20} (2005) 923, hep-th/0406003.

\bibitem{I}
 M.~B. Halpern, ``The orbifolds of permutation-type as physical
string systems at multiples of $c=26$: I. Extended actions and new
twisted world-sheet gravities,"{\em J. High Energy Phys.}  {\bf 0706} 
(2007) 068, hep-th/0703044.

\bibitem{II}
M.~B.Halpern, ``The orbifolds of permutation-type as physical string systems
at multiples of $c=26$:  II. The twisted BRST systems of $\hat c=52$ matter,'' 
{\em Int. J. Mod. Phys.} {\bf A22} (2007) 4587, hep-th/0703208.

\bibitem{III}
M.~B.Halpern, `` The orbifolds of permutation-type as physical string 
systems at multiples of $c=26$: III. The spectra of $\hat c=52$ strings,'' 
{\em Nucl. Phys.} {\bf B786} (2007) 297, hep-th/0704.1540.

\bibitem{IV}
M.~B.Halpern, ``The orbifolds of permutation-type as physical string 
systems at multiples of $c=26$: IV. Orientation orbifolds include orientifolds,''
{\em Phys. Rev.} {\bf D76} (2007) 026004, hep-th/0704.3667.

\bibitem{DV2}
R.~Dijkgraaf, E.~Verlinde and H.~Verlinde, ``Matrix string theory,''
  {\em Nucl. Phys.} {\bf B500} (1997) 43, hep-th/9703030.
  
\bibitem{Thorn}
C.~B. Thorn, ``Embryonic dual model for pions and fermions,'' {\em Phys. Rev.} 
{\bf D4} (1971) 1112.

\bibitem{Sew}
K.~Bardakci, M.~B.Halpern and J.~Shapiro, ``Unitary closed loops in 
Reggeized Feynman theory,'' {\em Phys. Rev.} {\bf 185} (1969) 1910; D.~Amati, C.~Bouchiat
and J.~L. Gervais,'' On the building of dual diagrams from unitarity,'' 
{\em Nuovo Cimento Lett.} {\bf2} (1969) 399; J.~Shapiro, ``Loop graph in 
the dual tube model,'' {\em Phys. Rev.} {\bf D5} (1972) 1945. 

\bibitem{Bant}
P.~Bantay, ``Characters and modular properties of permutation 
orbifolds,'' {\em Phys. Lett.} {\bf B419} (1998) 175, hep-th/9708120.

\bibitem{Cristo}
G.~Cristofano, V.~Marotta and G.~Niccoli, ``A new rational conformal 
field theory extension of the full degenerate $W_{1+\infty}^{(m)}$,'' 
hep-th/0412085.

\bibitem{Jank}
M.~Jankiewicz and T.~Kephardt, ``Extension of monster moonshine to 
$c=24k$ conformal field theories,'' hep-th/050178.

	
\end{thebibliography}
\end{document}